\documentclass[aps, showpacs, pra,   eqsecnum, twocolumn] {revtex4-1}
\usepackage{amsmath }
\usepackage{amssymb}
\newcommand*{\be}{\begin{equation}}
\newcommand*{\ee}{\end{equation}}
\newcommand*{\bea}{\begin{eqnarray}}
\newcommand*{\eea}{\end{eqnarray}}

\usepackage{graphicx}
\graphicspath{{pict/}{}}

 \DeclareFontFamily{OT1}{pzc}{}
 \DeclareFontShape{OT1}{pzc}{m}{it}%
 {<->  s  *  [1.400]  pzcmi7t}{}
\DeclareMathAlphabet{\mathscr}{OT1}{pzc}%
{m}{it}

\newcommand{\sech}{\, \mathrm{sech}}

\begin{document}

\title[]{Breathers in  $\mathcal{PT}$-symmetric optical couplers}

\author{I.  V.  Barashenkov$^{1,2,3}$, Sergey V. Suchkov$^{1,4}$, Andrey  A.  Sukhorukov$^1$,  Sergey  V. Dmitriev$^4$,  and Yuri S. Kivshar$^1$}
 \affiliation{$^1$ 
 Nonlinear Physics Centre, Research School of Physics and Engineering, Australian National University, Canberra ACT 0200,  Australia
 \\
 $^2$ Department of Mathematics, University of Cape Town, Rondebosch 7701, South Africa \\
 $^3$ Joint Institute for Nuclear Research, Dubna, Russia\\
 $^4$  Institute for Metal Superplasticity Problems, Russian Academy of Sciences, Ufa 450001, Russia
 }

\begin{abstract}
We show that the parity-time ($\cal{PT}$) symmetric coupled optical waveguides with gain and loss
support  localised  oscillatory structures  similar to the breathers of the classical $\phi^4$ model.
The power carried by the ${\cal PT}$-breather oscillates periodically, switching back and forth between the waveguides,
so that the gain and loss are compensated on the average.
The  breathers are found to coexist with solitons and be prevalent in the products of the soliton collisions.
We demonstrate that the evolution of the small-amplitude breather's envelope is governed
by a system of two coupled nonlinear Schr\"odinger equations, and employ this Hamiltonian
  system to show that the small-amplitude  ${\cal PT}$-breathers are stable.
 \end{abstract}

%

\maketitle

\section{Introduction}

Light propagation in $\mathcal{PT}$-symmetric optical systems
with balanced gain and loss has been under intense scrutiny in the past few years. The concept
has its roots in
quantum mechanics where a ${\cal PT}$ symmetric non-Hermitian
Hamiltonian
may have an entirely real spectrum of eigenvalues \cite{Bender:1998-5243:PRL,Bender:2007-947:RPP}.
 In optics, the $\mathcal{PT}$ symmetry  can be achieved by an appropriate
  modulation of the complex refractive index
\cite{Ruschhaupt:2005-L171:JPA, El-Ganainy:2007-2632:OL, Klaiman:2008-80402:PRL}.

The symmetric optical systems should display a variety of unusual
and often counter-intuitive phenomena
including  an unconventional beam refraction~\cite{Makris:2008-103904:PRL, Zheng:2010-10103:PRA}, Bragg scattering~\cite{Berry:2008-244007:JPA, Longhi:2010-22102:PRA},
 nonreciprocal Bloch oscillations~\cite{Longhi:2009-123601:PRL},
 symmetry-breaking transitions~\cite{Bendix:2009-30402:PRL, Ruter:2010-192:NPHYS},
a loss-induced optical transparency~\cite{Guo:2009-93902:PRL},
the conical diffraction~\cite{Ramezani:2012-13818:PRA}, a new type of Fano resonance~\cite{Miroshnichenko:2011-12123:PRA},
chaos~\cite{West:2010-54102:PRL}, and nonlocality manifested in the nontrivial effect of the boundaries
 \cite{Sukhorukov:2012-2148:OL}.
 Recently, optical $\mathcal{PT}$-symmetric couplers
 \cite{Guo:2009-93902:PRL,Ruter:2010-192:NPHYS} and lattices \cite{Regensburger:2012-167:NAT}
  have been realised experimentally.

Nonlinear effects in $\mathcal{PT}$-symmetric systems are of particular interest for
the fundamental  and applied science.
They offer potential for an efficient control of light, including the all-optical low-threshold switching~\cite{Chen:1992-239:IQE, Ramezani:2010-43803:PRA, Sukhorukov:2010-43818:PRA}
and unidirectional invisibility~\cite{Ramezani:2010-43803:PRA}.
In addition, nonlinearity can compensate the diffraction of stationary light beams and dispersion of  light pulses allowing the formation of spatial and temporal solitons. 


There has already been a large number of studies of optical solitons in $\mathcal{PT}$-symmetric systems.
Solitons in complex one-dimensional potentials were analyzed on the basis of the nonlinear Schr\"odinger equation
\cite{Musslimani:2008-30402:PRL, Dmitriev:2010-2976:OL, Hu:2011-43818:PRA, Abdullaev:2011-41805:PRA, Shi:2011-53855:PRA, Zhu:2011-2680:OL, Zezyulin:2011-64003:EPL, Nixon:2012-23822:PRA, Hu:2012-43826:PRA, Zezyulin:2012-43840:PRA, He:2012-3320:OC}.
The two-dimensional symmetric
potentials were dealt with in
  Refs.~\cite{Shi:2011-53855:PRA, Nixon:2012-23822:PRA, Zeng:2012-47601:PRE}.
The authors of \cite{Suchkov:2011-46609:PRE, Driben:2011-4323:OL, Driben:2011-51001:EPL, Alexeeva:2012-63837:PRA} classified
solitons in the planar $\mathcal{PT}$-symmetric couplers, whose geometry is  intermediate  between one- and two-dimensional lattices.

The $\mathcal{PT}$-symmetric solitons considered in the above publications
represented stationary self-localised modes. The solitons arise due to the exact compensation of  the gain and loss at each moment of time.
A more general type of localised objects was identified in \cite{Alexeeva:2012-63837:PRA} where the unstable solitons were observed to seed spatially-localised temporally-periodic states. (In the context of planar stationary waveguides, these are interpreted as the transversally localised structures with profiles oscillating along the waveguide.) These objects resemble breathers in conservative systems (such as the $\phi^4$ and sine-Gordon equation)~\cite{Kosevich:1974-1793:ZETF, *Dashen:1975-3424:PRD, *Segur:1987-747:PRL, *Boyd:1990-177:NLN}; hence they were referred to simply as {\it breathers\/}  \cite{Alexeeva:2012-63837:PRA}.

 In this paper the $\mathcal{PT}$ breathers are studied in more detail.
     First, we derive the amplitude equations for the oscillatory solutions in the planar $\mathcal{PT}$-symmetric nonlinear optical coupler
          (equations for the  envelopes of the oscillatory wavepackets).
     The amplitude equations turn out to be Hamiltonian --- despite the fact that the original system includes both gain and loss.
     These Hamiltonian equations are then  used
   to show that the (zero-velocity) ${\cal PT}$ breathers form two-parameter families
      with variable amplitude, localisation width, and contrast of power density oscillations.
      We  also employ these
    equations to establish the stability of the breathers with small amplitude.
    Finally,   the  planar  $\mathcal{PT}$-symmetric coupler is simulated numerically.
        Results of  our numerical simulations 
       demonstrate that the breathers are generic objects which are commonly formed as a result of  the soliton collisions.

The outline of the paper is  as follows. In Sec.~\ref{model}, we
introduce the mathematical model, and in the subsequent section,
derive  equations for the slowly-varying envelopes of its oscillatory solutions.
 Section \ref{Breather} uses these amplitude equations
 to classify the ${\cal PT}$-symmetric breather states.
 The stability of the small-amplitude breathers is established in section \ref{Stability}.
  In Sec.~\ref{Numerics}, we describe the formation of breathers in the soliton-soliton
  collisions.
Finally, Sec.~\ref{Conclusions} summarises results of this study.

\section{Model}
\label{model}

The $\mathcal{PT}$-symmetric coupler,
i.e.,
a pair of coupled waveguides with power gain in one waveguide and optical loss
of equal rate in the other, has been
studied theoretically \cite{El-Ganainy:2007-2632:OL, Klaiman:2008-80402:PRL, Ramezani:2010-43803:PRA, Sukhorukov:2010-43818:PRA, Li:2011-66608:PRE}
and experimentally \cite{Guo:2009-93902:PRL, Ruter:2010-192:NPHYS}. Optical systems that include the
$\mathcal{PT}$-symmetric coupler as a structural element \cite{Miroshnichenko:2011-12123:PRA, Dmitriev:2011-13833:PRA, Sukhorukov:2012-2148:OL}
and systems consisting of arrays of such couplers
 \cite{Dmitriev:2010-2976:OL, Zheng:2010-10103:PRA, Li:2011-66608:PRE, Szameit:2011-21806:PRA, Suchkov:2011-46609:PRE, Ramezani:2012-13818:PRA, Suchkov:2012-33825:PRA} have also been discussed in literature.

\begin{figure}[t]
 \begin{center}
    \includegraphics*[width=\linewidth]{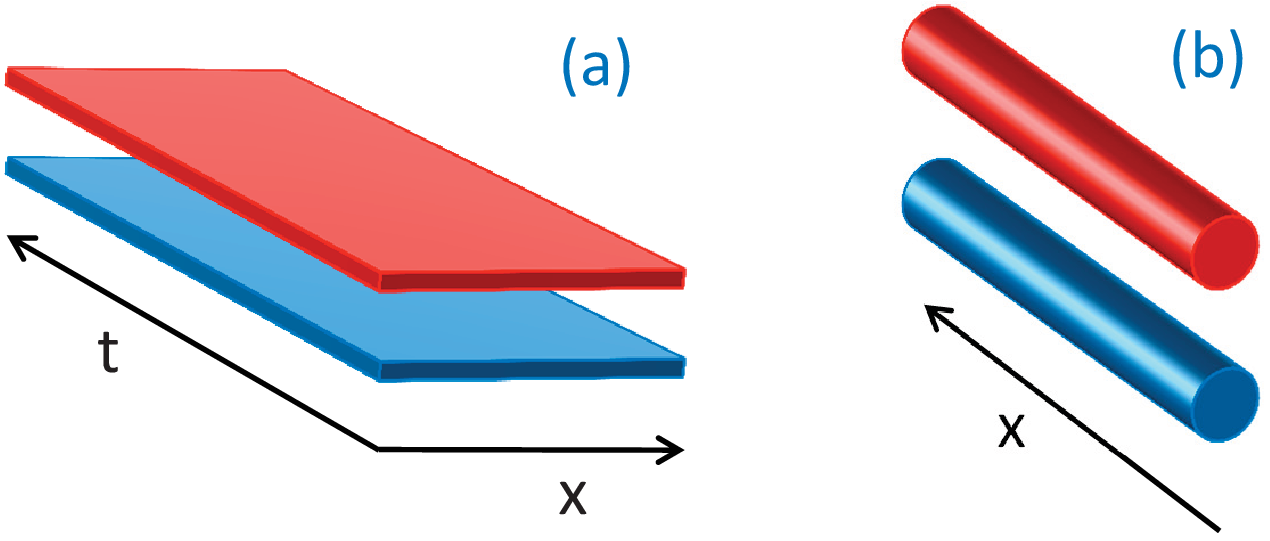}
            \caption{(Color online)
    A schematic
    representation of ${\cal PT}$-symmetric coupled waveguides with gain (red) and loss (blue waveguide).
(a)~Two planar waveguides carrying stationary light beams. Here $t$ and $x$
indicate the longitudinal  and transversal spatial coordinate, respectively.
(b)~A pair of one-dimensional waveguides  where light pulses  undergo temporal evolution as they travel along the $x$ axis.
   \label{Coupler}}
 \end{center}
 \end{figure}

Following \cite{Driben:2011-4323:OL, Alexeeva:2012-63837:PRA} we analyze the diffraction of optical beams propagating in a {\it planar\/}  $\mathcal{PT}$ coupler,
in media with the Kerr-type nonlinearity.  The amplitudes of the
 active and passive modes  in this setting satisfy a system of two coupled
 nonlinear Schr\"odinger equations,
\begin{equation} \label{A1}
 \begin{split}
    i u_t +u_{xx} + 2 |u|^2 u = -v + i \gamma u, \\
    iv_t+ v_{xx} + 2|v|^2v    = -u - i \gamma v.
 \end{split}
\end{equation}
Here $t$  is the (spatial) coordinate in the
propagation direction and $x$ is the transversal coordinate.
The  coefficient $\gamma>0$ is the amplification rate for the  waveguide with gain and, at the same time, the damping rate for the waveguide
with loss. This planar coupler is schematically shown in Fig.~\ref{Coupler}(a).
It is fitting to note here that the system \eqref{A1} emerges  as the  continuum limit of
 the chain of ${\cal PT}$
 couplers considered in \cite{Suchkov:2011-46609:PRE}.

The same $\mathcal{PT}$ symmetric system  \eqref{A1} can describe the propagation of  optical pulses (rather than stationary light beams)
\cite{Alexeeva:2012-63837:PRA}. This alternative interpretation of  Eqs.\eqref{A1} arises if
   $t$  and $x$ stand for the time and distance in the
 frame of reference travelling along with the pulse.
This is the arrangement illustrated by Fig.~\ref{Coupler}(b).

The system \eqref{A1} is not conservative.
Neither the individual powers associated with the two modes,
\be
\mathcal P_u=\int |u|^2 dx, \quad \mathcal P_v = \int |v|^2 dx,
\label{PP}
\ee
 nor their sum are conserved.
The total power satisfies
\be
\frac{d}{dt} (\mathcal{P}_u+\mathcal{P}_v)  = 2 \gamma (\mathcal{P}_u-\mathcal{P}_v),
\label{P}
\ee
which implies that it remains constant only on solutions which have ${\mathcal P}_u={\mathcal P}_v$
for all times \cite{Alexeeva:2012-63837:PRA}.

\section{Weakly nonlinear amplitude equations}
\label{AE}

We start our analysis by transforming  Eqs.~\eqref{A1} to a system with a diagonal linear part.
Assuming $\gamma < 1$ and defining
\be
a= \frac{e^{i \theta} u - v}{2 \omega_0},
\quad
b= \frac{e^{-i \theta} u+ v}{2 \omega_0},
 \label{B2}
\ee
where
\[
 \theta = \arcsin \gamma, \quad \omega_0= \cos \theta,
\]
Eqs.~\eqref{A1} are taken to
\begin{equation} \begin{split} 
ia_t +a_{xx}- \omega_0 a
+2(|a|^2 + 2 |b|^2) a \\
+ 4i e^{-i \theta} \gamma a^2 b^* + 2 e^{2 i \theta} a^* b^2=0,  \\
i b_t + b_{xx} + \omega_0 b + 2(2 |a|^2+ |b|^2) b \\ 
- 4i e^{i \theta} \gamma a^* b^2 + 2 e^{-2 i \theta} a^2 b^*=0.
\end{split}
\label{A30}
\end{equation}

The system \eqref{A30} has two simple reductions
or, equivalently, two invariant manifolds.
Letting $b=0$, Eqs.~\eqref{A30} reduce to
 a scalar  nonlinear Schr\"odinger equation
\be
ia_t +a_{xx}- \omega_0 a
+2|a|^2 a =0,  \label{F1}
\ee
while letting $a=0$ yields  a scalar Schr\"odinger equation
with the opposite sign of the frequency term:
\be
i b_t + b_{xx} + \omega_0 b + 2 |b|^2 b =0.  \label{F2}
\ee
Both \eqref{F1} and \eqref{F2} have soliton solutions and hence
the system \eqref{A30} admits two types of `simple' solitons:
one with $b=0$ and the other one with $a=0$.
These low- and high-frequency solitons have been analysed before
\cite{Suchkov:2011-46609:PRE,Driben:2011-4323:OL,Alexeeva:2012-63837:PRA}.
 Here, our aim is
to construct more general solutions with both components nonzero.

To this end, we note that when $a$ and $b$ are so small that the nonlinear
part in \eqref{A30} can be neglected, the resulting linear system has
a family of spatially homogeneous stationary-wave solutions: $a={\mathcal A}_0e^{-i \omega_0t}$,
$b={\mathcal B}_0e^{i \omega_0 t}$. To search for the nonlinear counterparts of
these, we consider a long-wavelength small-amplitude configuration:
\be
a(x,t)= \epsilon^{1/2}A(X,t),
\quad
b(x,t)= \epsilon^{1/2}B(X,t),
 \label{A100}
\ee
where $X= \epsilon^{1/2} x$ and  $\epsilon$ a small parameter   ($\epsilon>0$).
The $O(1)$ fields $A$ and $B$ satisfy
\begin{equation}
\label{A17}
  \begin{split} 
iA_t +\epsilon A_{XX}- \omega_0 A
+2 \epsilon (|A|^2 + 2 |B|^2) A \\ 
+ 4i e^{-i \theta} \epsilon \gamma A^2 B^* + 2 e^{2 i \theta} \epsilon  A^* B^2=0,  \\ 
i B_t + \epsilon B_{XX} + \omega_0 B + 2 \epsilon (2 |A|^2+ |B|^2) B  \\ 
- 4i e^{i \theta} \epsilon \gamma A^* B^2 + 2 e^{-2 i \theta}  \epsilon A^2 B^*=0.
\end{split} 
\end{equation} 

Solutions of the system \eqref{A17} can be sought for as expansions
 in powers of  $\epsilon$:
\be
A= A_0+ \epsilon A_1 + ..., \quad
B= B_0 + \epsilon B_1 + ....
\label{A102}
 \ee
We also assume   that the coefficients $A_n$ and $B_n$ depend on a hierarchy of `slow times' and `zoomed out'  spatial
coordinates: $A_n=A_n(T_0, T_1, ...; X_0, X_1, ...)$, $B_n=B_n(T_0, T_1, ...; X_0, X_1, ...)$, where
\be
T_n =\epsilon^n t, \quad X_n = \epsilon^n X, \quad n=0,1,2...
\label{B4}
 \ee
 In the limit $\epsilon \to 0$ the scaled time and space variables
decouple, and can be treated as independent.
In what follows, we adopt a shorthand notation
\[
D_n= \partial/\partial T_n, \quad
\partial_n = \partial/ \partial X_n.
\]

Note that the parameter $\epsilon$ is not pegged to any scale of the original model
\eqref{A1},\eqref{A30}. Therefore we expect it to be absorbable in the parameters of
solutions that we will end up with.

Substituting the expansions \eqref{A102} in \eqref{A17}, we equate coefficients of like powers of $\epsilon$. The order $\epsilon^0$ gives
\begin{align*}
(iD_0 - \omega_0)A_0=0, \\
(i D_0 + \omega_0)  B_0 =0,
\end{align*}
whence
\be
A_0= e^{-i \tau} p, \quad
B_0= e^{i \tau} q,
\label{A101}
 \ee
 with
 \[
 \tau= \omega_0 \, T_0.
 \]
 The coefficients $p$ and $q$ are functions of all variables except $T_0$.

The order $\epsilon^1$ produces
\begin{align}
(i D_0- \omega_0) A_1=
-[iD_1 A_0 + \partial_0^2 A_0
+ 2 (|A_0|^2 + 2 |B_0|^2) A_0 \nonumber \\
+ 4i e^{-i \theta} \gamma A_0^2 B_0^* + 2 e^{2 i \theta} B_0^2 A_0^*],   \nonumber \\ 
(iD_0+ \omega_0) B_1 = -[
iD_1B_0 + \partial_0^2 B_0
 + 2(|B_0|^2 + 2 |A_0|^2) B_0 \nonumber \\
 - 4i e^{i \theta} \gamma B_0^2 A_0^*
+ 2e^{-2 i \theta} A_0^2B_0^*].
\label{A31}
\end{align} 
To eliminate the secular terms, we impose
\begin{equation} \begin{split} 
iD_1p + \partial_0^2 p+ 2(|p|^2+ 2|q|^2) p=0, \\ 
iD_1q+ \partial_0^2 q + 2(|q|^2+ 2 |p|^2) q=0. \label{A19}
\end{split} \end{equation} 
The remaining terms in the right-hand sides of \eqref{A31} involve the third harmonics only; hence we get, for $A_1$ and $B_1$,
 \begin{equation} \begin{split} 
A_1=\frac{1}{2 \omega_0}
\left(e^{2i \theta} q^2 p^*e^{3i \tau} \right.
 \left. - 4i e^{-i \theta} \gamma p^2 q^* e^{-3i \tau}\right), \\ 
B_1= -\frac{1}{2 \omega_0}   \left(4i e^{i \theta} \gamma q^2 p^* e^{3i \tau} \right.
   \left. + e^{-2 i \theta} p^2 q^* e^{-3i \tau}\right).    \label{B6}
   \end{split} \end{equation} 

Proceeding  to the order $\epsilon^2$, and setting the
corresponding secular terms to zero, we obtain
\begin{equation} \begin{split} 
iD_2p + 2 \partial_0\partial_1 p+ \frac{1}{\omega_0}(|q|^2-2|p|^2)
|q|^2 p=0, \\ 
iD_2 q + 2 \partial_0 \partial_1 q+ \frac{1}{\omega_0}
(2|q|^2-|p|^2)|p|^2q=0,
 \label{A20}
 \end{split} \end{equation} 
where we have substituted for $A_1$ and $B_1$ from \eqref{B6}.

According to Eqs.\eqref{A19}, the variations in the amplitudes $p$ and $q$
become noticeable only over long periods of time, $\Delta t \sim \epsilon^{-1}$.
Eqs.\eqref{A20}
govern the evolution of these amplitudes over even
 longer time intervals, $\Delta t \sim \epsilon^{-2}$.
It is convenient to combine Eqs.\eqref{A19} and \eqref{A20} into a system that takes care of the evolution 
on {\it both\/} slow scales. 
To this end,  we add Eqs.\eqref{A19} 
to Eqs.\eqref{A20} multiplied by $\epsilon$
and define $T=\epsilon t$.
Since the amplitudes do not depend on $T_0$, the chain rule gives
$\partial/\partial T=
D_1+ \epsilon D_2+\epsilon^2 D_3+ ... \/$.
Thus, to within ${\mathcal O}(\epsilon^2)$, we have
$D_1 p+ \epsilon D_2p=p_T$ 
and $D_1 q+ \epsilon D_2q=q_T$,
and so the resulting pair of equations can be written as
\begin{align}
i   p_T +  p_{XX} +2 (|p|^2+ 2 |q|^2)p
+\frac{\epsilon}{\omega_0} (|q|^2-2 |p|^2) |q|^2 p=0,  \nonumber \\
i   q_T +  q_{XX} +2 (|q|^2+ 2 |p|^2)q
+\frac{\epsilon}{\omega_0} (2|q|^2-|p|^2) |p|^2 q=0. \label{A21}
\end{align} 
(Here  $\epsilon \geq 0$).
This is a hamiltonian system, with the Hamilton functional
\begin{eqnarray*}
H= \int \left[ |p_X |^2   + | q_X|^2
                               - (|p|^4+ |q|^4
        + 4 |p q|^2)          \right.       \nonumber \\  \left.
+ \epsilon \omega_0^{-1} |p q|^2 (|p|^2-|q|^2)
\right] dX.
\end{eqnarray*}

The amplitude equations \eqref{A21} describe the evolution of the 
slowly changing envelope  of a small-amplitude, weakly
localised packet of waves with the carrier frequency $\omega_0$.
Over time intervals $\epsilon^{-1} \lesssim \Delta t \lesssim \epsilon^{-2}$, equations \eqref{A21} are equivalent to the original system \eqref{A1}.
This remarkable equivalence of a dissipative and conservative system, holding for a particular but fairly broad class of trajectories,
is attributable to the ${\cal PT}$-symmetry of the former.

Setting $\epsilon= 0$,  the system \eqref{A21} becomes
 \begin{equation} \begin{split} 
i   p_T +  p_{XX} +2 (|p|^2+ 2 |q|^2)p =0, \\ 
i   q_T +  q_{XX} +2 (|q|^2+ 2 |p|^2)q =0.
 \label{A210} \end{split} \end{equation} 
This vector nonlinear Schr\"odinger equation has been extensively studied in literature
\cite{Kivshar:2003:OpticalSolitons,Ueda:1990-563:PRA,Malomed:1991-1388:OL,Mesentsev:1992-1497:OL,Kaup:1993-3049:PRE,Haelterman:1994-3376:PRE,Haelterman:1993-145:OC,Haelterman:1994-3376:PRE,Silberberg:1995-246:OL,Yang:1996-111:STAM,Yang:1997-92:PD,Yang:1997-61:STAM,Yang:2001-26607:PRE,Tan:2001-56616:PRE}.
On the other
hand, the system \eqref{A21} with $\epsilon \neq 0$ does not seem to have been discussed before.

Note that both Eq.\eqref{A21} and the ``curtailed" system \eqref{A210} govern 
the {\it small-amplitude\/} breathers only, 
with $u, v \sim \epsilon^{1/2}$. 
However Eq.\eqref{A21} has an advantage   over Eq.\eqref{A210}
in  that the former system has a longer range of validity. While Eq.\eqref{A210} ceases to be 
valid for times exceeding $\epsilon^{-1}$, Eq.\eqref{A21} remains accurate for times as long
as $\epsilon^{-2}$.

Another reason for  the evaluation of the second order corrections in the perturbation expansion,
is related to the conservativity of the amplitude equations \eqref{A21}  and  \eqref{A210}.
Once the first-order amplitude equations are found to be given by a hamiltonian system
[the system \eqref{A19}],  the question arises  whether this property 
is specific to the first-order evolution only. The fact that the second-order dynamics are also governed
by a hamiltonian system, suggests then that the conservativity is an inherent property of
the small-amplitude oscillations.
We conjecture  that this property is valid to all orders 
in the perturbation theory (and may only be violated by terms that lie
beyond all orders).

\section{Breather solutions}
\label{Breather}

 \begin{figure}
    \includegraphics*[width=\linewidth]{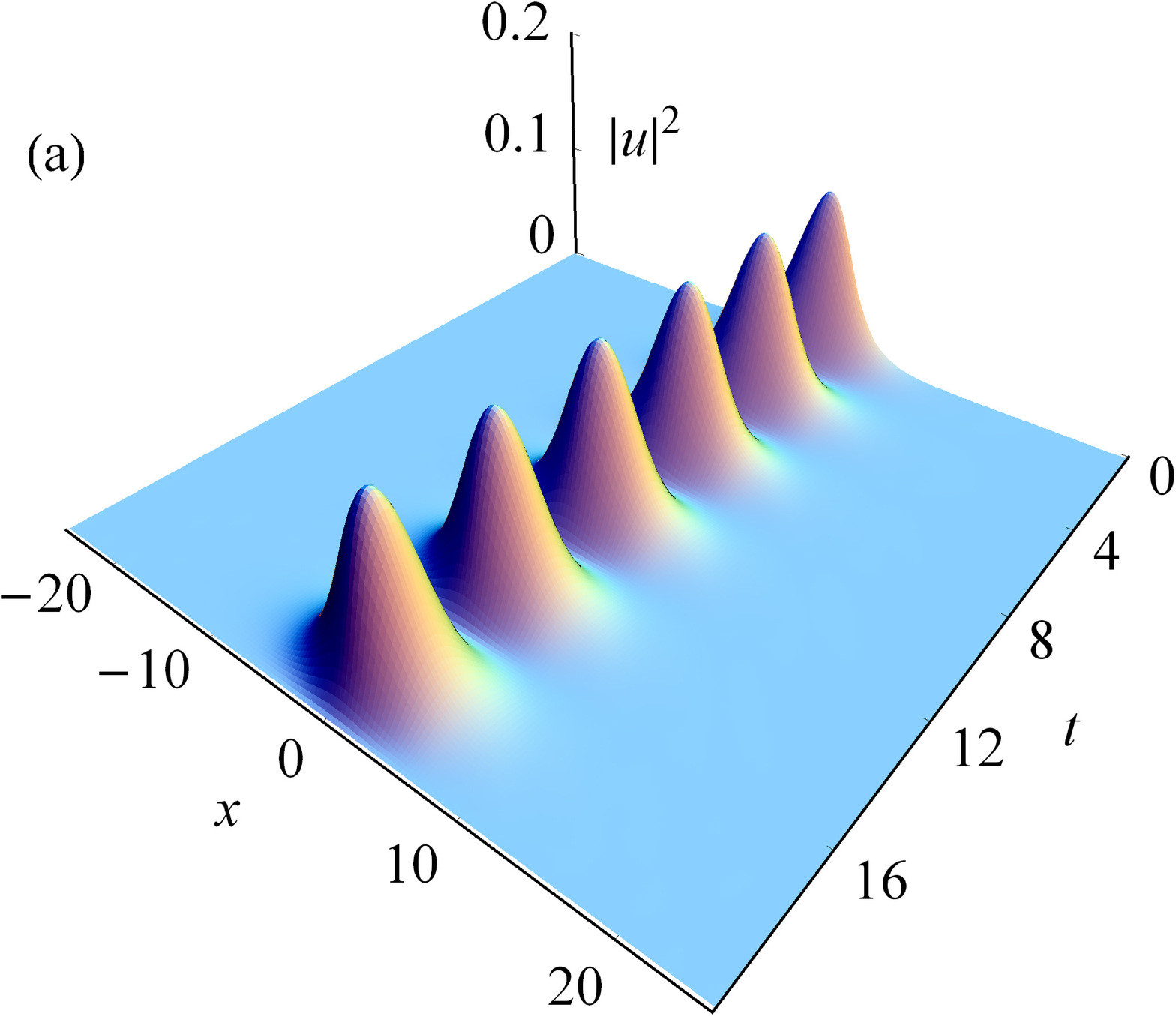}
    \includegraphics*[width=\linewidth]{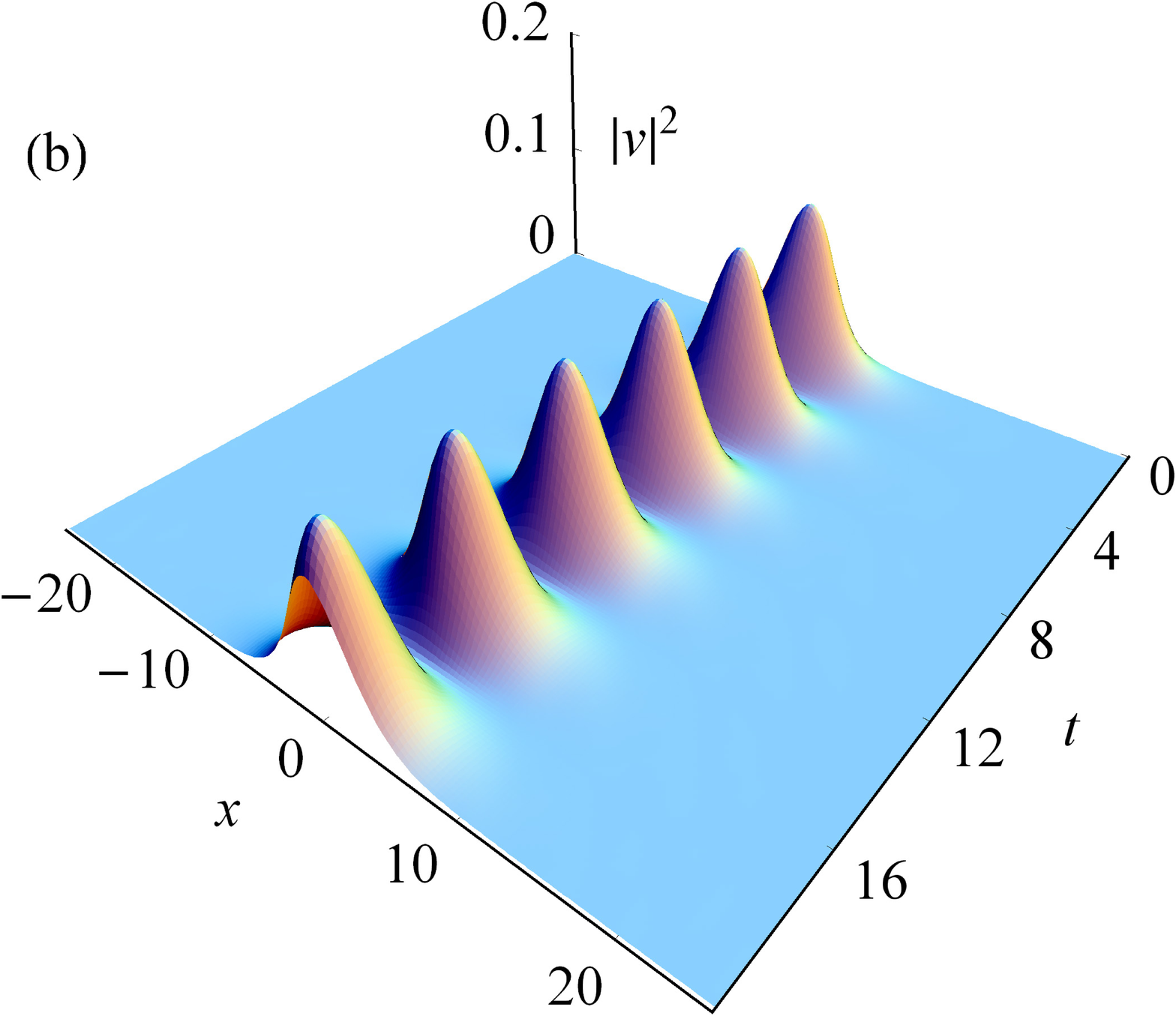}
     \includegraphics*[width=\linewidth]{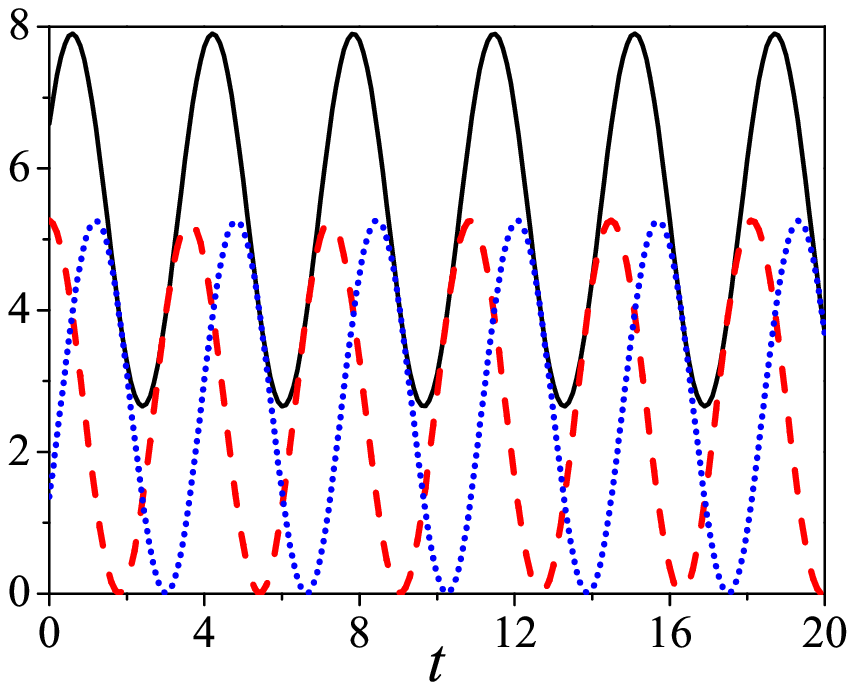}     \caption{(Color online) 
           Numerical evolution of the  initial condition in the form of the expansion
\eqref{A110}-\eqref{A105} with $t=0$.
(In this simulation,  $\gamma=0.5$ and $\epsilon=0.1$.)
Shown are $|u|^2$ (a), $|v|^2$ (b), and
 powers carried by the two components of the breather (c).
 In (c),  ${\mathcal P}_u$ is depicted by broken red and
 ${\mathcal P}_v$ by dotted blue line. Also shown is the total power
 ${\mathcal P}_u+ {\mathcal P}_v$ (solid line).
 The simulation continued until times much longer than $\epsilon^{-2}=100$,
 without any visible change in the amplitude or period of the breather.
 \label{breather_g99_C1_eps001}}
  \end{figure}

One simple solution of Eqs.~\eqref{A21} is
\be
p= e^{ i  \left( \mu T + \frac{V}{2} X\right)} \sqrt{\mu}
\sech [ \sqrt{\mu} (X- V T) ], \quad q=0.
\label{F3}
\ee
The other one is given by
\be
p=0, \quad
q= e^{ i \left( \nu T + \frac{W}{2} X \right)}
\sqrt{\nu} \sech [\sqrt{\nu} (X- W T)].
\label{F4}
\ee
These two solutions of \eqref{A21} will be referred to as {\it degenerate solitons}.
The parameters $\mu>0$, $\nu>0$, $V$ and $W$  can be chosen arbitrarily.
Here $\mu$ and $\nu$ give the amplitudes
of the degenerate solitons, and  $V$, $W$ are their velocities.

The degenerate soliton solutions of Eq.~\eqref{A21}  correspond to the solitons of the scalar reductions
\eqref{F1} and \eqref{F2} of the original system \eqref{A1}.
The degenerate soliton \eqref{F3} corresponds to the low-frequency soliton
of \eqref{A1}, and the solution
\eqref{F4} to its high-frequency counterpart \cite{Suchkov:2011-46609:PRE,Driben:2011-4323:OL,Alexeeva:2012-63837:PRA}.
The vector of the  power densities $\{ |u|^2, |v|^2 \}$ associated with each of these solutions
describes a pulse travelling, without oscillations, at the velocity $v=\epsilon^{1/2} V$ and $w=\epsilon^{1/2} W$, respectively.

Our main interest is in solutions of the system \eqref{A21}
which have both components nonzero. Thanks to
the Galilian invariance of \eqref{A21}, it is sufficient to consider separable
solutions corresponding to nonpropagating waves:
\be
p= e^{ i   \mu T} P(X), \quad
q= e^{i \nu T}  Q(X).
\label{G2}
\ee
The spatial parts  $P$ and $Q$ satisfy
\begin{align}
  P''- \mu P  +2 (|P|^2+ 2 |Q|^2)P
+\frac{\epsilon}{\omega_0} (|Q|^4-2 |P Q|^2) P=0, \nonumber  \\
 Q'' -\nu Q +2 (|Q|^2+ 2 |P|^2)Q
 +\frac{\epsilon}{ \omega_0} (2|P Q|^2-|P|^4)  Q=0,
  \label{A22}
\end{align}
where we use the notation $'' = d^2/d X^2$.

Localised solutions of  the stationary equations \eqref{A22}  give rise to oscillatory, breather-like,
configurations in the original model \eqref{A1}:
\begin{align*}
u(x,t)= \epsilon^{1/2} \left[  Qe^{i \omega_2 t}
+ Pe^{-i \omega_1 t} \right]+O(\epsilon^{3/2}), \\
v(x,t)= \epsilon^{1/2} \left[Qe^{i (\omega_2 t+\theta) }
- Pe^{-i (\omega_1 t+ \theta) }\right]+O(\epsilon^{3/2}),
\end{align*}
where $P=P(\epsilon^{1/2} x)$, $Q=Q(\epsilon^{1/2} x)$, and
\[
\omega_1= \omega_0- \epsilon \mu,
\quad
\omega_2=\omega_0+ \epsilon \nu.
\]
The corresponding $|u|^2$ and $|v|^2$ are
\begin{align*}
|u|^2=\epsilon \left( |P|^2 +  |Q|^2 \right)+ \epsilon \left[ Q P^*
e^{i(\omega_1+ \omega_2)t} + c.c. \right], \\
|v|^2=\epsilon \left( |P|^2 +  |Q|^2 \right)   - \epsilon \left[
Q P^*
e^{i(\omega_1+ \omega_2)t+ 2 i \theta} + c.c. \right],
\end{align*}
where $c.c.$ stands for the complex conjugate of the immediately preceding term
and we neglected the $O(\epsilon^2)$-corrections.
These quantities show temporal oscillations with the frequency $\omega_1+\omega_2=2 \cos \theta + \epsilon(\nu-\mu)$.

In this paper, we confine ourselves to  the simplest choice of $\nu=\mu$.
(A brief comment on
a more general situation with $\nu \neq \mu$ is in the Appendix \ref{MG}.)
An additional simplification is attained by  restricting to real solutions.
For real $P$ and $Q$ equations \eqref{A22}
  reduce to
\begin{align}
  P''- \mu P  +2 P^3+ 4 Q^2 P
+\frac{\epsilon}{\omega_0} (Q^2 -2 P^2) Q^2 P=0,  \nonumber  \\
 Q'' -\mu Q +2 Q^3+ 4 P^2 Q
 +\frac{\epsilon}{ \omega_0} (2 Q^2-P^2)P^2  Q=0. \label{A24}
\end{align} 

When $\epsilon=0$,  the system \eqref{A24} has an explicit solution
\be
P_0(X)= Q_0(X)= \sqrt{\frac{\mu}{3}} \sech ( \sqrt{\mu} X).
\label{C1}
\ee
The terms proportional to $\epsilon$ in \eqref{A24} are regular perturbations,
i.e.,  the perturbed solution
satisfying the boundary conditions
$P(X), Q(X) \to 0$ as $ |X| \to \infty$
exists for all sufficiently small $\epsilon$.
To show this,  we expand $P$ and $Q$
in powers of $\epsilon$,
 \be
 P= P_0+ \epsilon P_1+ \epsilon^2 P_2 +...,
\quad
Q = Q_0 + \epsilon Q_1 + \epsilon^2 Q_2 +...,
 \label{G1}
 \ee
and substitute the expansions  in \eqref{A24}.
Letting $\mathcal S=P_1+Q_1$
and $\mathcal D=Q_1-P_1$, the order $\epsilon$ gives
\begin{align}
(-d^2/d\xi^2 + 1 - 6 \sech^2 \xi) \,  \mathcal S= 0, \label{A25} \\
\left(- \frac{d^2}{d \xi^2} + 1 - \frac23 \sech^2 \xi \right) \mathcal D=  \frac{2}{9\sqrt{3}}
\frac{\mu^{3/2}}{ \omega_0} \sech^5 \xi,
 \label{A26}
\end{align}
where we have  defined $\xi=\mu^{1/2} X$.

The operator in the left-hand side of \eqref{A25} has a zero eigenvalue,
with the associated eigenfunction being odd. If we wish to construct a
solution with  definite parity (i.e. an even solution), we should take $\mathcal S=0$.
On the other hand,
the operator in the left-hand side of \eqref{A26} is positive definite,
hence invertible. As a result, Eq.~\eqref{A26} has an exponentially decaying solution:
\[
\mathcal D= \frac{1}{51 \sqrt{3}} \frac{\mu^{3/2}}{\omega_0} ( 6 \sech \xi+ \sech^3 \xi).
\]
Taken together with ${\mathcal S}=0$, this implies
\be
Q_1=-P_1= \frac{1}{102 \sqrt{3}} \frac{\mu^{3/2}}{\omega_0} ( 6 \sech \xi+ \sech^3 \xi).
\label{P1Q1}
\ee


Returning to the original variables $u$ and $v$ we  note that, as expected,  the
parameters $\epsilon$ and $\mu$ enter the solution only in combination $\epsilon \mu$.
Without loss of generality,  we can set one of these to 1,
e.g. $\mu=1$.

Equations \eqref{G2}, with $P$ and $Q$ expanded as in  \eqref{G1},
and  $P_n$, $Q_n$ as in \eqref{C1}, \eqref{P1Q1}
provide solutions to the amplitude equations \eqref{A21}:
\begin{align}
p=\frac{e^{i  T}}{\sqrt{3}}
\sech X \left[
1- \frac{\epsilon}{102 \omega_0} (6+ \sech^2 X) + O(\epsilon^2) \right],  \nonumber \\
q=\frac{e^{i  T}}{\sqrt{3}}
\sech X \left[
1+ \frac{\epsilon}{102 \omega_0} (6+ \sech^2 X) +O(\epsilon^2)  \right].
\label{B200}
\end{align} 
Since both $p$ and $q$ are nonzero in \eqref{B200}, we will be referring to these
solutions as {\it two-component} solitons.

Feeding Eqs.~\eqref{B200}   in
\eqref{A100}, \eqref{A102},
 \eqref{A101}, \eqref{B6}
gives
\begin{equation} \begin{split} 
a= \epsilon^{1/2} \left[ A_0+ \epsilon A_1+ O(\epsilon^2) \right],   \\  
b= \epsilon^{1/2} \left[B_0+ \epsilon B_1+ O(\epsilon^2) \right],
\label{A110} \end{split} \end{equation} 
with
\begin{widetext}
\begin{equation} \begin{split} 
A_0= \frac{e^{-i(\omega_0-\epsilon)t}}{\sqrt{3}} \sech (\epsilon^{1/2} x) \left[1 -\frac{\epsilon}{102 \omega_0} \left(6+ \sech^2 (\epsilon^{1/2} x)\right)+ O(\epsilon^2) \right], \\
B_0= \frac{e^{i(\omega_0+\epsilon)t}}{\sqrt{3}} \sech (\epsilon^{1/2} x) \left[1 + \frac{\epsilon}{102 \omega_0} \left(6+ \sech^2 (\epsilon^{1/2} x)\right)+ O(\epsilon^2) \right], \\
A_1= \frac{e^{ i \epsilon t}}{6 \sqrt{3} \omega_0}\sech^3 (\epsilon^{1/2} x)
\left[ e^{i( 3 \omega_0 t+ 2 \theta)}-4 i \gamma e^{-i(3 \omega_0 t+ \theta)} \right]+ O(\epsilon), \\
B_1=- \frac{e^{ i \epsilon t}}{6 \sqrt{3} \omega_0}\sech^3 (\epsilon^{1/2} x)
\left[4i \gamma e^{i(3 \omega_0 t+ \theta)} + e^{-i(3 \omega_0t + 2 \theta)}
  \right]+ O(\epsilon).
  \label{A111}
 \end{split} 
  \end{equation}
\end{widetext}
Equations \eqref{A110}-\eqref{A111}, taken together with the
 conversion formulas
\be
u(x,t)=a+b, \quad
v(x,t)=e^{i \theta} b- e^{-i \theta} a,
\label{A105}
\ee
yield solutions of the original equation \eqref{A1}.

To test the accuracy of the asymptotic solution \eqref{A110}-\eqref{A105},
we simulated equations \eqref{A1} with the initial conditions 
in the form \eqref{A110}-\eqref{A105} with
 $t=0$. [In these initial conditions, we neglected the $O(\epsilon^2)$ terms in $A_0, B_0$
and the $O(\epsilon)$ terms in $A_1, B_1$.] The resulting oscillatory configuration is
plotted in Fig. \ref{breather_g99_C1_eps001}. The fundamental harmonic in the frequency spectrum
of  $|u|^2$ and $|v|^2$ was indeed found to be very close to $2 \omega_0$,
the double frequency of the asymptotic solution.

As
we mentioned in section \ref{model}, the  system \eqref{A1} may be thought of as a continuum limit of
a chain of coupled $\mathcal{PT}$-symmetric dimers. The power in each dimer can
perform a periodic  oscillation \cite{Ramezani:2010-43803:PRA,Sukhorukov:2010-43818:PRA},
with an amplitude-dependent period.
The breather is an oscillation involving the entire chain. Although the
amplitude of oscillation varies along the chain, the coupling synchronises individual
dimers so that the breather has a single base frequency.
Accordingly, the power integrals    \eqref{PP} associated with the two modes show a
perfectly
periodic behaviour [Fig.~\ref{breather_g99_C1_eps001}(c)].

The total power ${\mathcal P}_u+ {\mathcal P}_v$ is not a constant of motion
but is periodic and therefore,  conserved on average.

 \section{Stability}
 \label{Stability}

The amplitude equations \eqref{A21} may be used to study
the dynamics of  the  solitons and  breathers  of the original system
\eqref{A1} over times up to $t \sim \epsilon^{-2}$. In particular, Eqs.~\eqref{A21}
may be used to study the stability of these objects.

Consider a stationary solution \eqref{G2}  of the system \eqref{A21}.
This can be one of the two degenerate solitons \eqref{F3} and \eqref{F4} ---
or the nondegenerate soliton \eqref{G1},\eqref{C1},\eqref{P1Q1} corresponding to the breather of the original
system \eqref{A1}.
We consider the simplest situation where $\mu=\nu$; in this case we may set, without loss of generality,
$\mu=\nu=1$.
Linearising Eqs.~\eqref{A21} about the stationary solution
and assuming perturbations of the form
\begin{align*}
\delta p(X,T)= e^{i T} \left[ \mathrm f(X,T)+i  \mathrm g(X,T)\right], \\
\delta q(X,T)= e^{i T} \left[ \mathrm y(X,T)+i \mathrm z(X,T)\right],
\end{align*}
where $\mathrm f, \mathrm g, \mathrm y$ and $\mathrm z$ are real, gives
\begin{equation} \begin{split} 
{ \mathcal L}_1 \mathrm f + \mathcal V (X) \mathrm y= - \mathrm g_T, &  \quad 
 \mathcal L_0 \mathrm g  =   \mathrm f_T,   \\  
{\mathcal M}_1 \mathrm y+ \mathcal V(X) \mathrm f = - \mathrm z_T,   &  \quad 
{\mathcal M}_0 \mathrm z    =   \mathrm  y_T.
\label{A106}
\end{split} \end{equation} 
Here we have introduced the operators
\begin{align*}
{\mathcal L}_0=-\partial^2/\partial X^2+ 1 -2 P^2-4Q^2 + \frac{\epsilon}{\omega_0} (2P^2-Q^2)Q^2,  \\
{\mathcal L}_1= -\partial^2/\partial X^2+ 1 - 6P^2-4Q^2+ \frac{\epsilon}{\omega_0} (6P^2-Q^2)Q^2, \\
{\mathcal M}_0=-\partial^2/\partial X^2+ 1 -4P^2-2Q^2 + \frac{\epsilon}{\omega_0} (P^2-2Q^2)P^2, \\
{\mathcal M}_1=  -\partial^2/\partial X^2+ 1 -4 P^2 -6Q^2 + \frac{\epsilon}{\omega_0} (P^2- 6Q^2)P^2,
\end{align*}
and a coefficient function
\[
\mathcal V(X)= -8PQ+ \frac{4 \epsilon}{\omega_0} (P^2-Q^2) PQ.
\]

For separable solutions of the form
\begin{align*}
\mathrm f(X,T)= \mathrm{Re} \, \left[ e^{\lambda   T} f(X) \right], \quad
\mathrm g(X,T)= \mathrm{Re} \, \left[ e^{\lambda T} g (X) \right],  \\
\mathrm y(X,T)= \mathrm{Re} \, \left[ e^{\lambda T} y(X) \right], \quad
\mathrm z(X,T)= \mathrm{Re} \, \left[ e^{\lambda  T} z(X)\right],
\end{align*}
with complex $f, g, y, z$, and $\lambda$, Eq.~\eqref{A106} reduces to an eigenvalue problem:
\be
{\mathscr A} \left( \begin{array}{c} {\vec y} \\ {\vec z} \end{array} \right) =
 \lambda
 \left( \begin{array}{c} {\vec y} \\ {\vec z} \end{array} \right),
 \label{A202}
 \ee
 where
 \be
 {\mathscr A}= \left(
 \begin{array}{cc}
 0 &  \mathcal H_0 \\
  - \mathcal H_1 & 0
  \end{array}
  \right)
  \label{A201}
  \ee
  is a $4 \times 4$ matrix with blocks given by
  \[
\mathcal H_0=
\left(
\begin{array}{cc}
\mathcal L_0  & 0 \\
0 & \mathcal M_0
\end{array}
\right),
\quad
\mathcal H_1=
\left(
\begin{array}{cc}
\mathcal L_1  & \mathcal V(X) \\
\mathcal V(X)  & \mathcal M_1
\end{array}
\right),
\]
and ${\vec y}$, ${\vec z}$ are two-component vectors:
  \be
  {\vec y}= \left( \begin{array}{c} f \\ y \end{array} \right),
  \quad
   {\vec z}= \left( \begin{array}{c} g  \\ z \end{array} \right).
   \label{A204}
   \ee

\subsection{Stability of the high- and low-frequency solitons}

Consider,
first, the degenerate soliton \eqref{F3} and let 
the velocity $V=0$.  [This degenerate soliton with $Q=0$ describes the amplitude
of the low-frequency soliton of the original ${\cal PT}$-symmetric equations \eqref{A1}.]
In this case,  the operators $\mathcal L_0$ and $\mathcal  L_1$ reduce to $L_0$ and $L_1$, respectively, where
\begin{align}
L_0= -d^2/dX^2+ 1 -2  \sech^2  X, \label{L0}   \\
L_1= -d^2/dX^2+ 1 -6  \sech^2 X, \label{Longhi:2009-123601:PRL}
\end{align}
while  $\mathcal M_0$ and $\mathcal M_1$ acquire  a common form which we denote $L_{\frac12}$:
\be
L_{\frac12}= -\frac{d^2}{dX^2}+ 1 -4  \sech^2 X
+ \frac{\epsilon}{\omega_0}  \sech^4  X.
\label{Longhi:2009-123601:PRL2}
\ee
Since $Q=0$ implies $\mathcal V(X)=0$,  the eigenvalue problem \eqref{A202}  acquires a block-diagonal form:
\begin{align}
L_0 g= \lambda f, \quad L_1 f =-\lambda g,  \label{NS} \\
L_{\frac12}  \, y= -\lambda z, \quad L_{\frac12} \, z = \lambda y. \label{ab}
\end{align}

Eq.~\eqref{NS} is the linearised eigenvalue problem for the scalar cubic
nonlinear Schr\"odinger equation, a well researched integrable system.
It has no discrete eigenvalues except the four-fold zero eigenvalue. Its continuous spectrum occupies
the imaginary axis.

On the other hand, Eq.~\eqref{ab} gives
\[
L_{\frac12}^2 y= -\lambda^2 y.
\]
 This implies that  $\lambda=i \omega$, where $\omega$
is an eigenvalue of the hermitian operator $L_{\frac12}$. Since all such eigenvalues are real,
all $\lambda$'s are pure imaginary and hence the degenerate soliton is stable.

When $\epsilon=0$, the operator $L_\frac12$   has two
discrete eigenvalues,  $\omega_a$ and $\omega_b$, given by
\be
\omega_a= \alpha-3 \approx    -1.438,
\quad
\omega_b =3\alpha-4 \approx    0.685,
\label{fr}
\ee
with $\alpha=(\sqrt{17}-1)/2$.
The corresponding eigenfunctions are  $\psi_a= \sech^\alpha X$ and $\psi_b=\sech^{\alpha-1} X \tanh X$, respectively.
The eigenvalues $\omega_a$ and $\omega_b$
persist when $\epsilon$ deviates from zero (but remains finitely small). It is only
when $\epsilon$ grows above a certain finite value that $\omega_b$ and then $\omega_a$ immerse in the continuous spectrum.
Accordingly,  for $\epsilon$ below a finite threshold,
the degenerate soliton \eqref{F3} has two modes of internal oscillation.
(For $\epsilon=0$, this fact has been established in \cite{Yang:1997-61:STAM}.)

 The degenerate soliton \eqref{F4}
 corresponds to the high-frequency soliton of the original equations \eqref{A1}.
The linearisation about this degenerate soliton
 leads to the
same eigenvalue problem \eqref{ab}, with the same operator
\eqref{Longhi:2009-123601:PRL2},  where one just needs to replace $\epsilon \to - \epsilon$.
This observation establishes the stability of the soliton \eqref{F4}.
As long as $\epsilon$ remains below a finite threshold, the operator $L_\frac12$ with $\epsilon \to - \epsilon$
has two discrete eigenvalues;
 hence the degenerate soliton \eqref{F4}  has two internal modes.

The fact that the degenerate solitons 
of the amplitude equations \eqref{A21} are stable implies that
both the low- and high-frequency solitons of the  ${\cal PT}$-symmetric
system \eqref{A1} are stable for sufficiently small $\epsilon$.
This conclusion is in agreement with the analysis  of the
low- and high-frequency soliton performed directly on
 the equations \eqref{A1}. Namely,
the high-frequency soliton was shown to be stable when its
amplitude $a$ lies below a finite threshold $a_c$, $a_c=\left(\frac23\right)^{1/2} (1-\gamma^2)^{1/4}$
\cite{Driben:2011-4323:OL,Alexeeva:2012-63837:PRA}.
On the other hand, the low-frequency soliton  has an unstable eigenvalue
irrespectively of the amplitude
but its  real part is exponentially small when the amplitude is small \cite{Alexeeva:2012-63837:PRA}.
This instability constitutes an effect that lies beyond all orders in $\epsilon^n$;
it  cannot be captured by the amplitude equations \eqref{A21}.
The  unstable perturbations take an exponentially long time to grow in this case;
hence  the small-amplitude low-frequency soliton will not reveal any instability 
when studied over time intervals $t \sim \epsilon^{-n}$.

The frequencies of the  internal modes of the low-  and high-frequency soliton
solutions of Eqs.~\eqref{A1} were also computed in \cite{Alexeeva:2012-63837:PRA}. These
coincide with the frequencies \eqref{fr}  computed using
the amplitude equations \eqref{A21}.

\subsection{Stability and spectrum of breather: $t \sim \epsilon^{-1}$}

Turning to the two-component
 soliton  \eqref{B200},
 it is convenient to  consider the soliton of the ``curtailed" system \eqref{A210}  first. 
  The stability  of the 
  soliton of the system \eqref{A210} will imply
   the stability  of 
  the   breather
 of the original $\mathcal{PT}$-symmetric system \eqref{A1} over time intervals $t \sim \epsilon^{-1}$
 (where $\epsilon^{1/2}$ is the amplitude of the breather).

The two-component  soliton  of  the system \eqref{A210} is given by  Eqs.~\eqref{B200} with $\epsilon = 0$:
\be
p= \frac{1}{\sqrt{3}}  e^{iT} \sech X, \quad
q= \frac{1}{\sqrt{3}}  e^{iT} \sech X.
\label{A211}
\ee
Depending on the context, this symmetric solution was referred to  as the
 {\it  linearly polarised\/} \cite{Haelterman:1994-3376:PRE} or
{\it equally mixed\/}       \cite{Yang:1996-111:STAM}   soliton.
Note that setting $\epsilon=0$ in Eqs.~\eqref{B200} does not mean that 
we are considering breathers of zero amplitude.   The  nonzero parameter $\epsilon$
 remains present in the corresponding breather solution 
  \eqref{A110}, \eqref{A111}, \eqref{A105};
  in particular the amplitude of the breather remains equal to $\epsilon^{1/2} \neq 0$.

The stability of the soliton \eqref{A211} was proved by the construction of a Lyapounov functional \cite{Mesentsev:1992-1497:OL}.
With an eye to addressing the situation of general $\epsilon$, we reconsider
the stability of this solution here --- using the eigenvalue analysis.

When $\epsilon=0$,  the eigenvalue problem \eqref{A202}
can be cast in the block-diagonal form
\begin{align}
\left( \begin{array}{cc}
0 & -L_1 \\ L_0 & 0
\end{array} \right) \left( \begin{array}{c} \zeta_1 \\ \zeta_2
\end{array} \right) = \lambda \left( \begin{array}{c} \zeta_1 \\ \zeta_2
\end{array} \right), \label{H2} \\
\left( \begin{array}{cc}
0 & -L_+ \\ L_0 & 0
\end{array} \right) \left( \begin{array}{c} \rho_1 \\  \rho_2
\end{array} \right) = \lambda \left( \begin{array}{c} \rho_1 \\ \rho_2
\end{array} \right), \label{H3}
\end{align}
where
the operators $L_0$ and $L_1$
are as in \eqref{L0}-\eqref{Longhi:2009-123601:PRL},  and
\be
L_+= -\frac{d^2}{dX^2} + 1 -\frac23  \sech^2 X.
\label{Lp}
\ee
The components of the column vectors in \eqref{H2}-\eqref{H3} are the sums and differences of the
components of the vectors in \eqref{A204}:
$\zeta_1=z+g$, $\zeta_2=y+f$,
$\rho_1=z-g$, $\rho_2=y-f$.

Eq.~\eqref{H2} arose in the previous section [see Eq.~\eqref{NS}].
It
is the linearised eigenvalue problem for the
scalar cubic nonlinear Schr\"odinger equation.
As discussed there, the matrix-differential operator \eqref{H2}  does not have any discrete eigenvalues except
the four zeros.
Therefore Eq.~\eqref{H2} can be safely disregarded and we can focus on Eq.~\eqref{H3}.

In order to transform  Eq.~\eqref{H3} to a form more amenable to analysis,
we note that
the only discrete eigenvalue of the operator \eqref{Lp} is $\beta+ 1/3$, where $\beta=\sqrt{11/12}- 1/2 >0$.
(It is associated with the nodeless eigenfunction $\psi=\sech^\beta X$.)  Hence the operator $L_+$ is
positive definite and admits an inverse.
This observation allows us to write  the vector equation \eqref{H3} as a generalised eigenvalue problem for
a pair of scalar operators,
\be
L_0 \rho_1 =  -\lambda^2 L_+^{-1} \rho_1. \label{H4}
\ee

In \eqref{H4},  $L_0$ is a symmetric operator, and $L_+^{-1}$ symmetric and positive definite.
All eigenvalues $(-\lambda^2)$ of
the problem  \eqref{H4} are real and the corresponding eigenfunctions can also be chosen real.
The lowest eigenvalue, $-\lambda_0^2$,   can be found as the minimum of the Rayleigh quotient:
\be
-\lambda_0^2 = \min \frac{(\rho_1, L_0 \rho_1)}{(\rho_1, L_+^{-1} \rho_1)}.
\label{H5}
\ee
Here $(  ,  )$ stands for the scalar product in the space of square integrable
real  functions: $(\phi, \psi)=\int_{-\infty}^\infty \phi(X) \psi(X) dX$.

The lowest eigenvalue of  the Schr\"odinger operator $L_0$
is zero; it is associated with the nodeless eigenfunction $z^{(0)}(X)= \sech X$.
Therefore the Rayleigh quotient in \eqref{H5} cannot take negative values and its minimum is
exactly zero:
 $-\lambda_0^2=0$. This means that the matrix-differential operator
 in the left-hand side of  \eqref{H3} does not have any nonzero real
eigenvalues $\lambda$ and so  the soliton \eqref{A211} of the vector nonlinear Schr\"odinger \eqref{A210} is stable.

This is the main conclusion of this subsection. It implies that the small-amplitude breather of the
$\mathcal{PT}$-symmetric system \eqref{A1} is stable over time intervals $t \sim \epsilon^{-1}$.

In fact it is not difficult to show that the operator \eqref{H3} does not have
any discrete eigenvalues at all --- neither real nor imaginary. (See the Appendix \ref{IM}.)
The implication is
 that when $\epsilon=0$, the two-component soliton  of the vector nonlinear Schr\"odinger 
does not have internal modes.
(This fact has been previously established by numerical means  \cite{Yang:1997-61:STAM}.)
With regard to the breather of the $\cal{PT}$-symmetric system \eqref{A1}, this implies
that the small-amplitude breather cannot have any modulating frequencies of order $\epsilon$  in its spectrum.
This is the second conclusion of this subsection.

\subsection{Stability of the breather: $t \sim \epsilon^{-2}$}

To extend the breather stability result to times of order $\epsilon^{-2}$, we need to
consider  the   system \eqref{A21}
with  $\epsilon \neq 0$.
We should demonstrate that its solution  \eqref{B200}
 does not have 
unstable eigenvalues  with $\mathrm{Re} \, \lambda$
 of order $\epsilon^\sigma$, $0< \sigma \leq 1$, 
in its spectrum.

We  begin the stability analysis 
of this solution with the  identification of  symmetries of the system \eqref{A21}.
These will provide information on zero eigenvalues of the operator \eqref{A201}.

Besides the translation and Galilean invariance,
the system  \eqref{A21} is
 symmetric with respect to the $U(1) \times U(1)$ transformations of the form $p \to p e^{i \phi}$, $q \to q e^{i \chi}$,
 where $\phi, \chi = \mathrm{const}$.
 In addition, $\mu$ and $\nu$
 can be chosen arbitrarily in the stationary system \eqref{A22}.
Thus each solution of the form \eqref{G2} is a member of a six-parameter
continuous family and therefore,
the eigenvalue problem \eqref{A202} has six zero eigenvalues.

The corresponding eigenvectors and generalised eigenvectors of the matrix $\mathscr A$  can be found explicitly.
First, we observe that
\be
\mathcal H_0 \left(\begin{array}{c} P \\ 0 \end{array} \right)=
\mathcal H_0 \left(\begin{array}{c} 0 \\  Q \end{array} \right)=0,
\label{zm}
\ee
and  $\mathcal{H}_1 (P_X, Q_X)^T=0$; hence
$(P,0,0,0)^T$, $(0,Q,0,0)^T$, and $(0,0,P_X,Q_X)^T$  are the $U(1)$ and translational
eigenvectors, respectively.  One can also check  that
$\mathcal{H}_1(P_\mu, Q_\mu)^T=-(P,0)^T$,
$\mathcal{H}_1(P_\nu, Q_\nu)^T=-(0,Q)^T$ and
$\mathcal H_0 {\vec w}=(P_X, Q_X)^T$, where ${\vec w}= -\frac12 X(P,Q)^T$.
These define  the generalised eigenvectors: $(0,0,P_\mu, Q_\mu)^T$, $(0,0, P_\nu, Q_\nu)^T$,  and  $-\frac12 X (P,Q,0,0)^T$.

All nonzero eigenvalues $\lambda$ of the matrix $\mathscr A$ can be found from
the solution of the eigenvalue problem for a $2 \times 2$ matrix:
\be
 \label{FEV}
\mathcal{H}_0
\mathcal{H} _1 \left( \begin{array}{c}
f \\ y \end{array} \right)= - \lambda^2 \left( \begin{array}{c}
f \\ y \end{array} \right).
\ee
Using \eqref{zm} one can readily check that
the eigenvectors of $\mathcal H_0 \mathcal H_1$  corresponding to  $-\lambda^2 \neq 0$
satisfy
\[
\int f(X) P(X) dX= \int y(X) Q(X) dX=0.
\]
These orthogonality constraints define a subspace of the space of square integrable vector-functions.
On this subspace, the operator $\mathcal H_0$ admits an inverse and \eqref{FEV} can be written as
\be
\mathcal{H}_1
\left( \begin{array}{c}
f \\ y \end{array} \right) = -\lambda^2                                 \mathcal H_0^{-1}  
\left( \begin{array}{c}
f \\ y \end{array} \right).
\label{H1}
\ee

The components $P(X)$ and $Q(X)$ of the solution \eqref{G1} remain
positive for all $X$ as long as $\epsilon$ remains small.
This means that  zero remains the lowest eigenvalue of
the
 operators $\mathcal L_0$ and $\mathcal M_0$ ---
 the operators whose null eigenvectors are given by $P$ and $Q$.
 Therefore, the operator ${\mathcal H}_0^{-1}$ remains positive  definite (and symmetric) --- while the operator $\mathcal H_1$ is
 symmetric.
 Eq.~\eqref{H1} implies then that all eigenvalues $(-\lambda^2)$ are real,
so that all  $\lambda$ are either real or pure imaginary.

As $\epsilon$ grows from zero, the six  eigenvalues
of the matrix $\mathscr A$  remain at the origin. New
discrete eigenvalues can only arise by
bifurcating from the continuous spectrum
which fills the imaginary axis of $\lambda$ outside the gap $(-i, + i )$.
Once an eigenvalue has detached from the continuum, it
can move along the imaginary axis toward the origin. 
However the eigenvalue could only reach the origin as $\epsilon$ exceeded a finite threshold.
 Therefore, the two-component soliton will remain stable as long as $\epsilon$
remains small.

Concerning the breather solution of the system \eqref{A1}, the implication of this
result is that the ${\cal PT}$-symmetric breather is stable on the timescale
$t \lesssim \epsilon^{-2}$.
 (That is, the breather's lifetime is no shorter than $\epsilon^{-2}$).

 \section{Breather production in soliton collisions}
 \label{Numerics}

Breathers  are known  not  to be
 exceptional or isolated  occurrences in the ${\cal PT}$-symmetric planar coupler.
 In particular, they  form as a result of the soliton instability \cite{Alexeeva:2012-63837:PRA,Driben:2012:EPL}.
 In this section we argue that breathers are even more common than solitons themselves:
 a collision of a high- and a low-frequency soliton produces two or more breathers,
 and a collision of two breathers also results in one or more of these oscillatory objects.

We use Eqs.~\eqref{A1} to simulate the evolution of the initial condition in the
form  of two  solitons
of equal amplitudes, moving toward each other with equal velocities:
\begin{equation} \begin{split} 
u(x,0)=a+b, \quad
v(x,0)=e^{i \theta}b - e^{-i \theta}a,  \\
a= e^{ i   \frac{v}{2} (x+x_0) } \sqrt{\mu}
\sech [ \sqrt{\mu} (x+x_0) ], \\
b= e^{ -i \frac{v}{2} (x-x_0)}
\sqrt{\mu} \sech [\sqrt{\mu} (x-x_0)].
\label{IC}
\end{split} \end{equation} 
Taking $x_0>0$,  the low-frequency soliton is initially on the left and
the high-frequency one is on the right; the initial
velocities are $v$ and $-v$, respectively.
[Note that  in \eqref{IC}, the same symbol
  $v$ denotes  the velocity of the soliton and the second component
 of the vector field, $v(x,t)$;  this
 slight abuse of notation should cause no confusion.]
The initial distance between the solitons is
assumed to be much larger than their widths: $2 \sqrt \mu x_0 \gg1$.

The high-frequency soliton is stable if $\mu \leq \frac23 \sqrt{1-\gamma^2}$ \cite{Driben:2011-4323:OL,Alexeeva:2012-63837:PRA}.
The low-frequency soliton is unstable for all $\mu$ but
when the amplitude is small, its instability
growth rate is exponentially small in $\mu$ \cite{Alexeeva:2012-63837:PRA}.
Therefore when the solitons' amplitudes are sufficiently small,
the low-frequency soliton will not manifest instability in
the run-up to the collision. The two small-amplitude solitons can be
considered as two stable entities.

The  collision of the low-frequency and the high-frequency solitons
in the ${\cal PT}$ symmetric system \eqref{A1} corresponds to the
collision of degenerate solitons \eqref{F3}-\eqref{F4} governed by the amplitude equations \eqref{A21}.
In the particular case $\epsilon =0$, such collisions were studied by Tan and Yang \cite{Tan:2001-56616:PRE}
(see also \cite{Yang:1996-111:STAM}).
Depending on the solitons' initial velocities, the colliding degenerate solitons were recorded to
pass through each other or bounce back.  The solitons emerging from the collision would no longer be degenerate;
instead, they would have both $p$ and $q$ components nonzero
  \cite{Yang:1996-111:STAM,Tan:2001-56616:PRE}.
  Translated in the language of the ${\cal PT}$-system \eqref{A1}, this means
  that the collision of the small-amplitude ${\cal PT}$ solitons should typically result in the emergence of
  two breathers.

\begin{figure}[t]
 \begin{center}
      \includegraphics*[width=\linewidth]{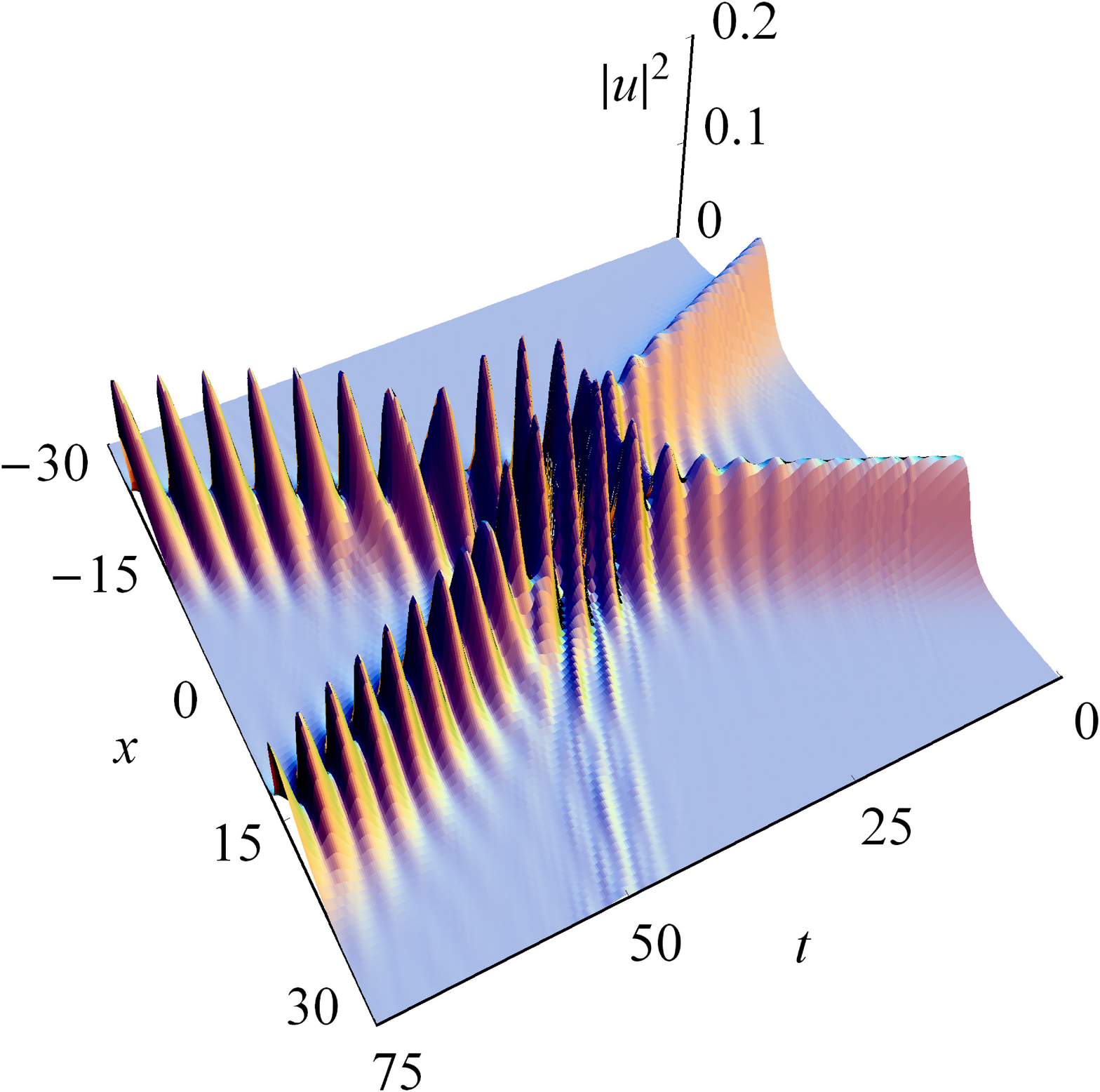}
   \caption{(Color online) 
   The collision of the low- (initially on the left)
  and the high-frequency soliton (initially on the right).
    As the solitons approach each other, they develop the beat-frequency
    oscillations of growing amplitude. The localised objects emerging
    from the collision remain oscillatory despite the growing separation distance
    ---
    these are a pair of breathers.
    The breathers are weakly radiating; also note the emission of a rapid
    small-amplitude
    breather at the moment of collision.
     In this simulation, $\gamma=0.5$,   $\sqrt{\mu}=0.3$, $v=0.4$, and $x_0=16$.
    \label{typical}}
 \end{center}
 \end{figure}

 This is indeed the scenario that we have observed in our numerical  simulations of Eqs.~\eqref{A1}.
 We have detected the formation of two breathers  in collisions of small- and moderate-amplitude solitons.
 A typical evolution is depicted in Fig.~\ref{typical}.

\begin{figure}
 \begin{center}
       \includegraphics*[width=\linewidth]{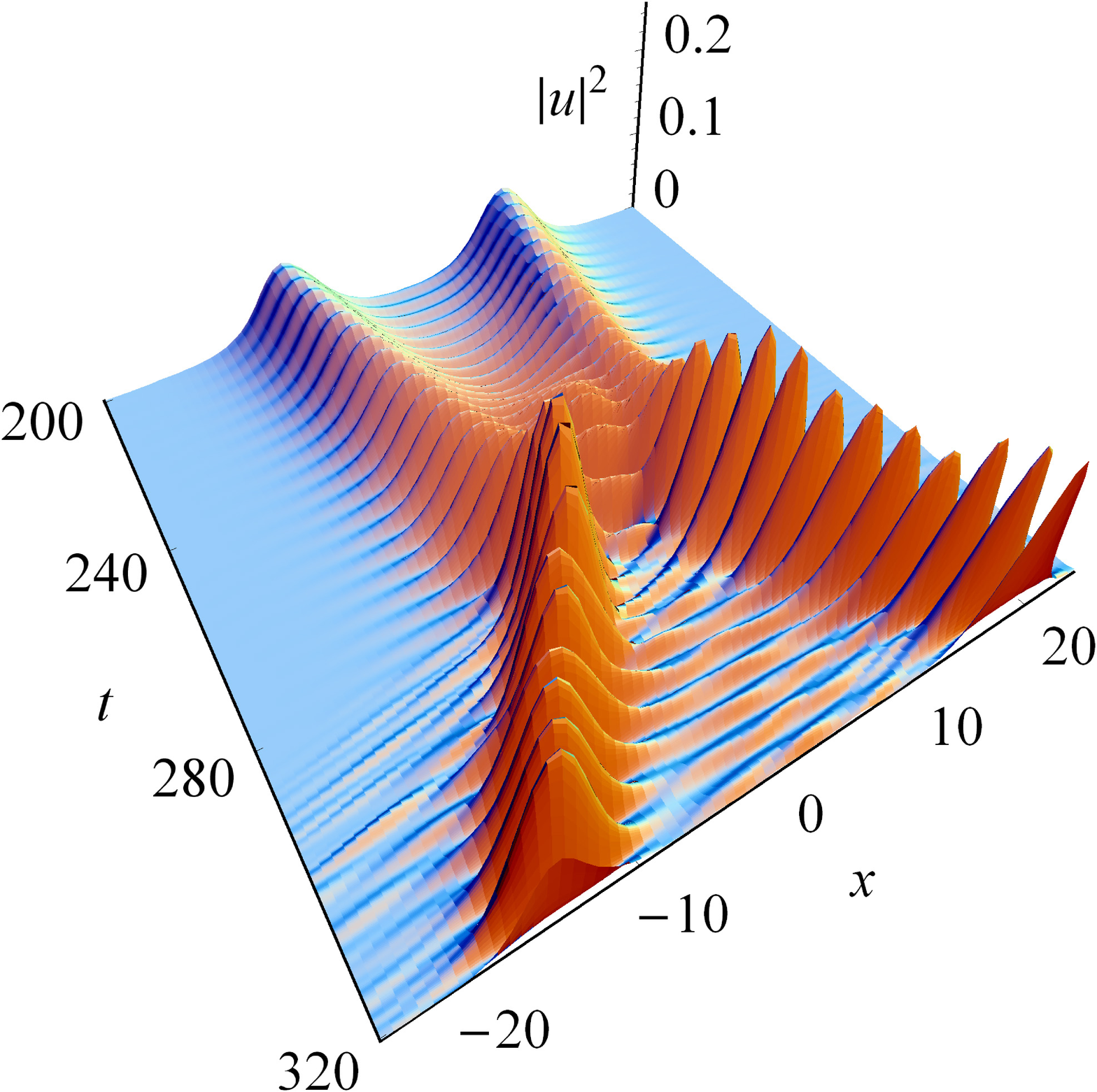}
       \includegraphics*[width=\linewidth]{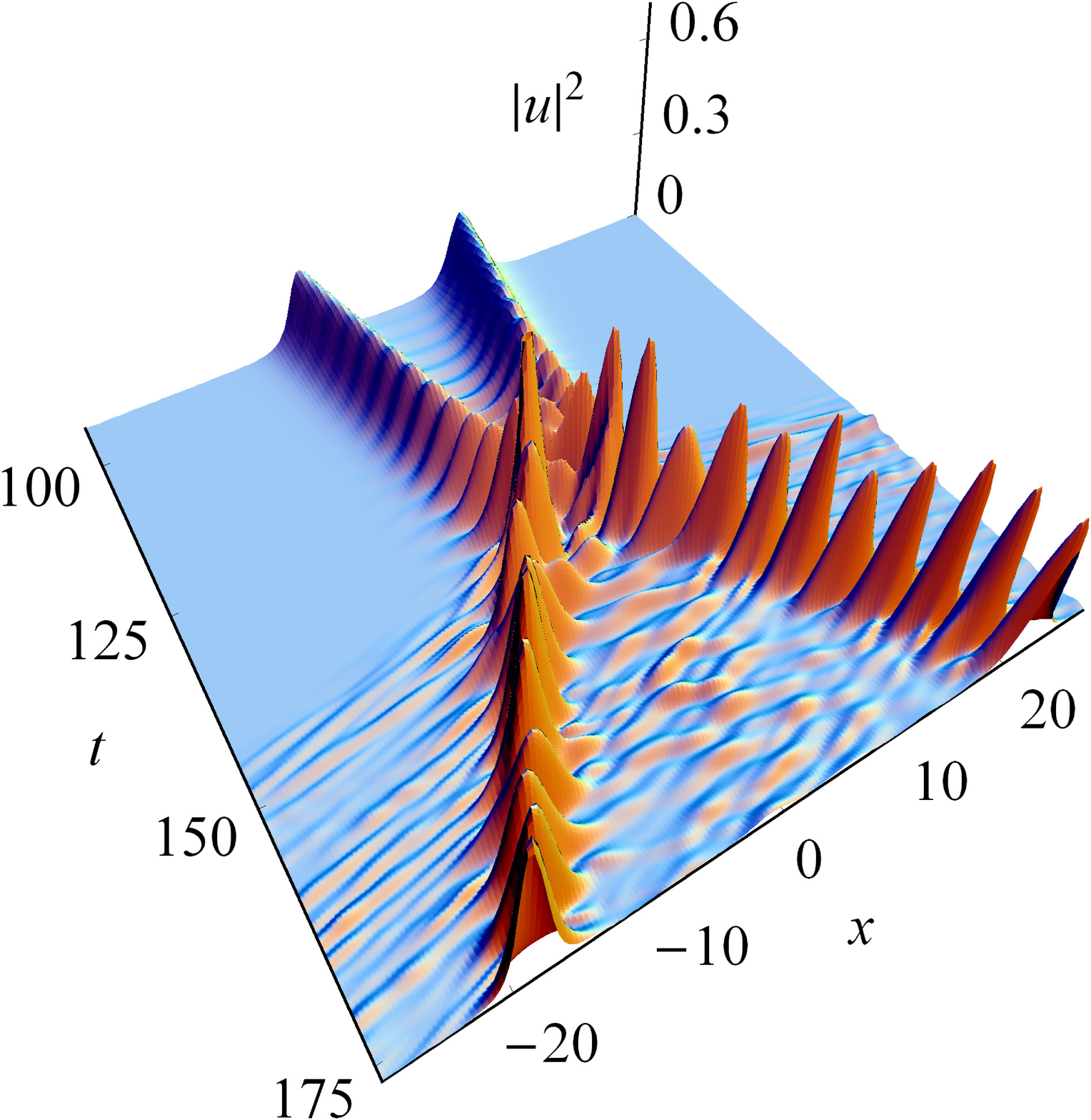}
         \includegraphics*[width=\linewidth]{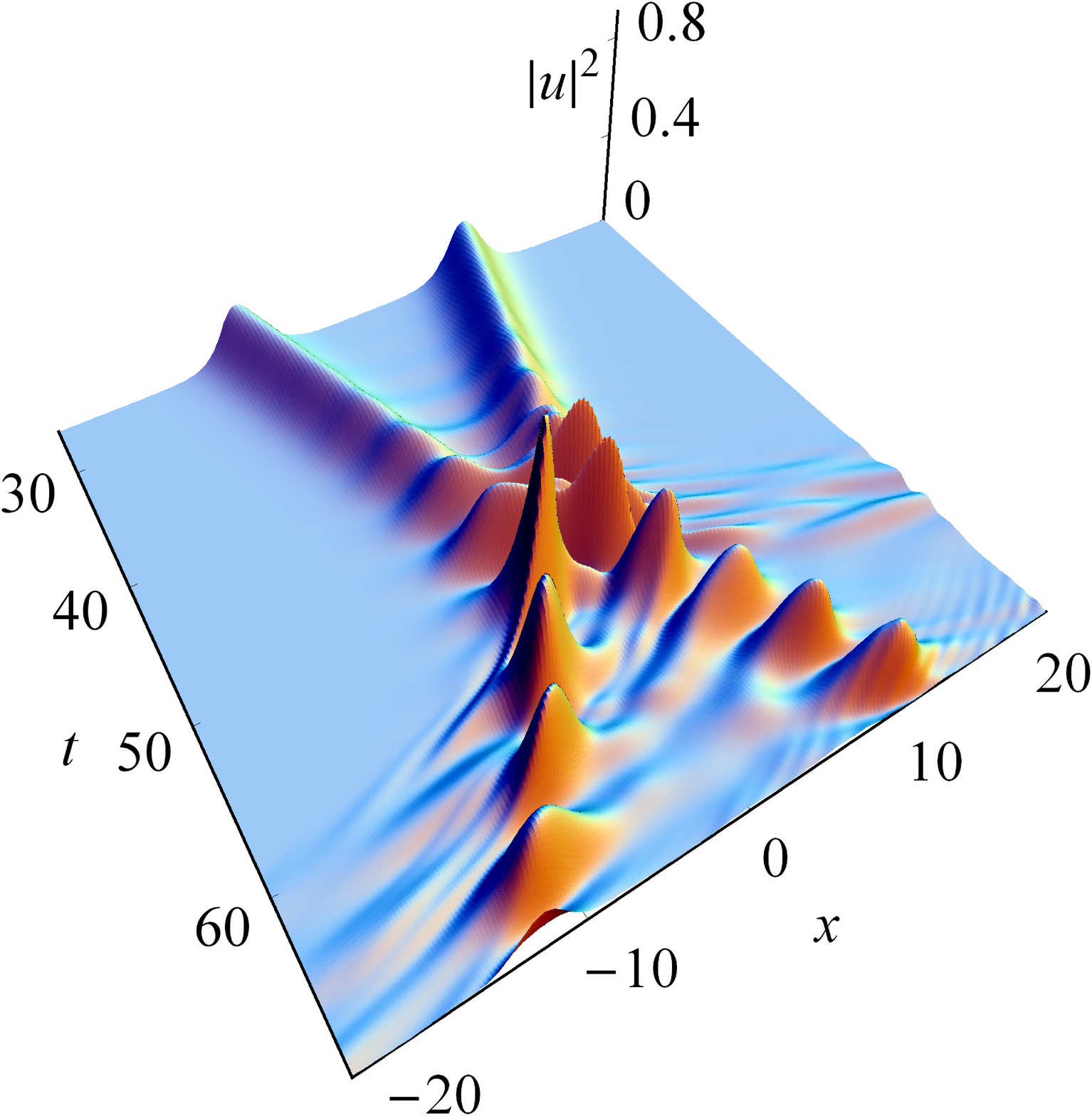}
                  \caption{(Color online) 
                  The collision of solitons with moderate
                          and large  amplitudes, small  and large initial velocities.
                          In  (a), $\gamma=0.6$, $\sqrt{\mu}=0.3$,  and the initial velocity
   $v=0.075$ 
 lies in the interval $(0.1 \mu^{1/2}, 0.3 \mu^{1/2})$.
  Note a small-amplitude nonpropagating breather left behind while
  two large-amplitude  fragments shoot out of the collision.
                   The panel  (b)     shows the collision of solitons with larger amplitudes.
 Here  $\gamma=0.5$, $\sqrt{\mu}=0.5$,  and $v=0.125$. The panel
 (c) corresponds to large initial velocities:
  $\gamma=0.5$, $\sqrt{\mu}= 0.25$, and $v=0.6$.
  \label{collisions}        }
 \end{center}
 \end{figure}

An interesting feature of the degenerate-soliton  collisions
recorded by Tan and Yang \cite{Tan:2001-56616:PRE}, was that
the reduction of the collision velocity would not result in the decrease of
the velocities of the solitons after collision.
In agreement with this amplitude-equation effect, our simulations of the
collision of  ${\cal PT}$ solitons with initial velocities $v \to 0$ have produced
breathers diverging at finite speeds (see e.g. Fig.~\ref{collisions}(a,b)).

Another  inelastic effect detected  in the  curtailed
  amplitude equation \eqref{A210},  pertained to the initial velocities in the  range $0.1< V <0.3$.
  For these $V$, the collision of two degenerate solitons   was seen    to
 result in  the production of a stationary small-amplitude soliton, in addition to the two transmitted or reflected ones    \cite{Tan:2001-56616:PRE}.
  A similar phenomenon accompanies
 the collision of the low- and high-frequency small-amplitude solitons in
  our
 ${\cal PT}$-symmetric system \eqref{A1}. Namely,
 the initial condition  \eqref{IC} with
 $v$ in the range $0.1 \mu^{1/2} < v < 0.3 \mu^{1/2}$ and small $\mu$
 gives rise to three breathers. Two of these
   move apart while the third, small-amplitude, breather is left behind near the origin.
   We have observed
  this effect even for not-very small soliton amplitudes, Fig. \ref{collisions}(a).

As the amplitudes of the colliding solitons are increased,
the curtailed equation \eqref{A210}  ceases to be applicable.
The collision of larger-amplitude solitons
is now accompanied by intense radiation,
while the oscillations of the emerging breathers acquire a low-frequency modulation
 [Fig.~\ref{collisions}(b)].
As the amplitudes exceed a certain threshold,
the collision results in a blowup of one of the fragments.

One more range of parameter values where
the equation \eqref{A210} does not furnish any accurate description of
the dynamics, pertains to large $v$.
As $v$ is increased, we observe the growth of the transient amplitude of
one of the emerging breathers  ---  a kind of a rogue wave
appearing just after the collision [Fig.~\ref{collisions}(c)].
Eventually, this rogue wave seeds the blow-up of the breather.

It is worth emphasising here that the creation of breathers is characteristic only
for the collision of two solitons of different types
(that is, collision of the low-  with the  high-frequency soliton).
The scattering of  two {\it like} solitons, e.g. two high-frequency solitons,  is purely elastic --- for the simple reason that
the initial condition and the resulting solution
belong to the same invariant manifold  $a=0$.
The  constraint  $a=0$ defines a reduction to a completely integrable
 equation [Eq.~\eqref{F2}], hence the elasticity of collisions.

The ubiquity of the breathers stems from the fact that they are not
confined to the $a=0$ or $b=0$ manifolds.
They represent trajectories evolving out of generic initial conditions
which do not belong to either of the two reductions.

\begin{figure}[t]
 \begin{center}
    \includegraphics*[width=\linewidth]{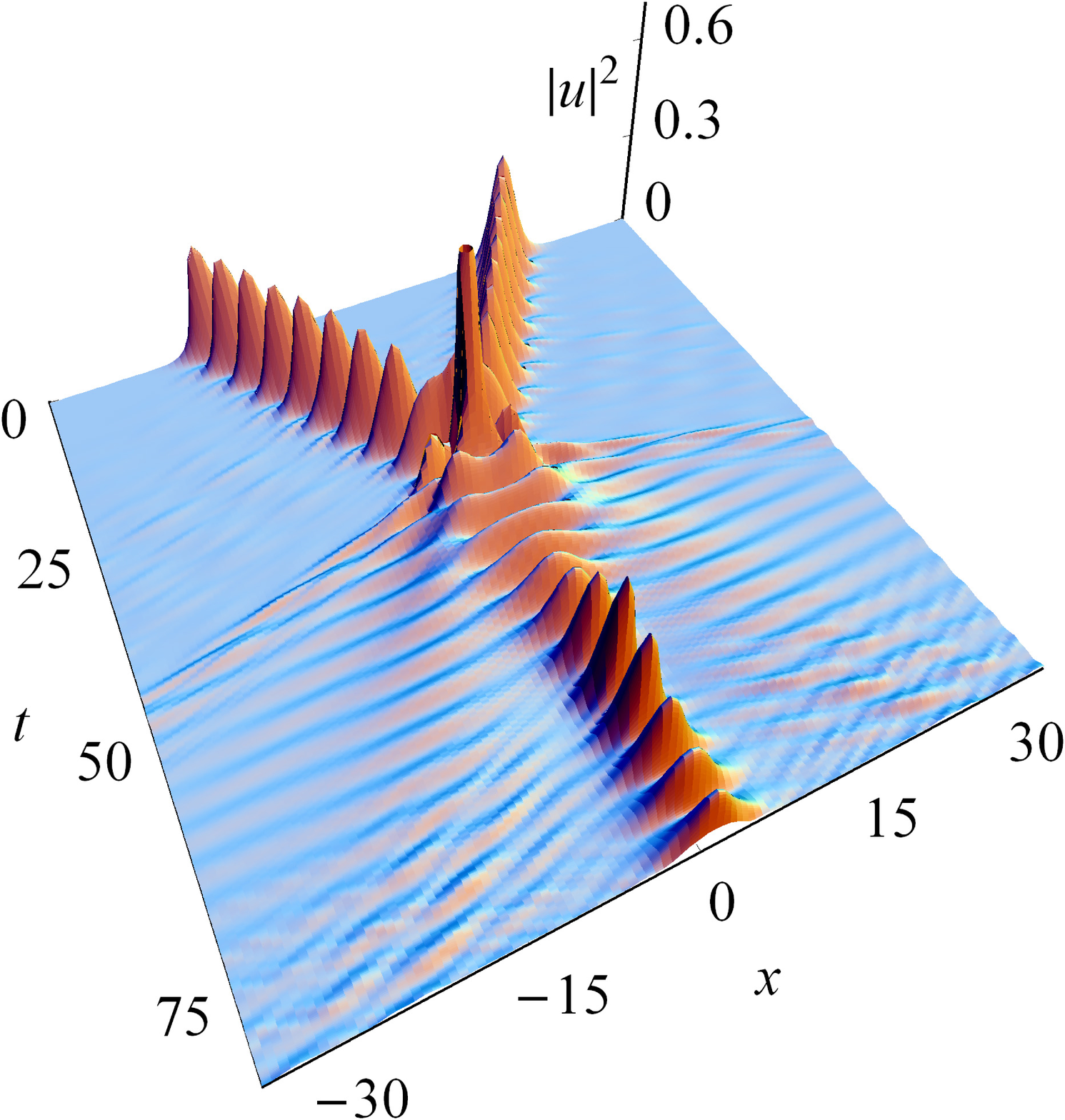}
   \includegraphics*[width=\linewidth]{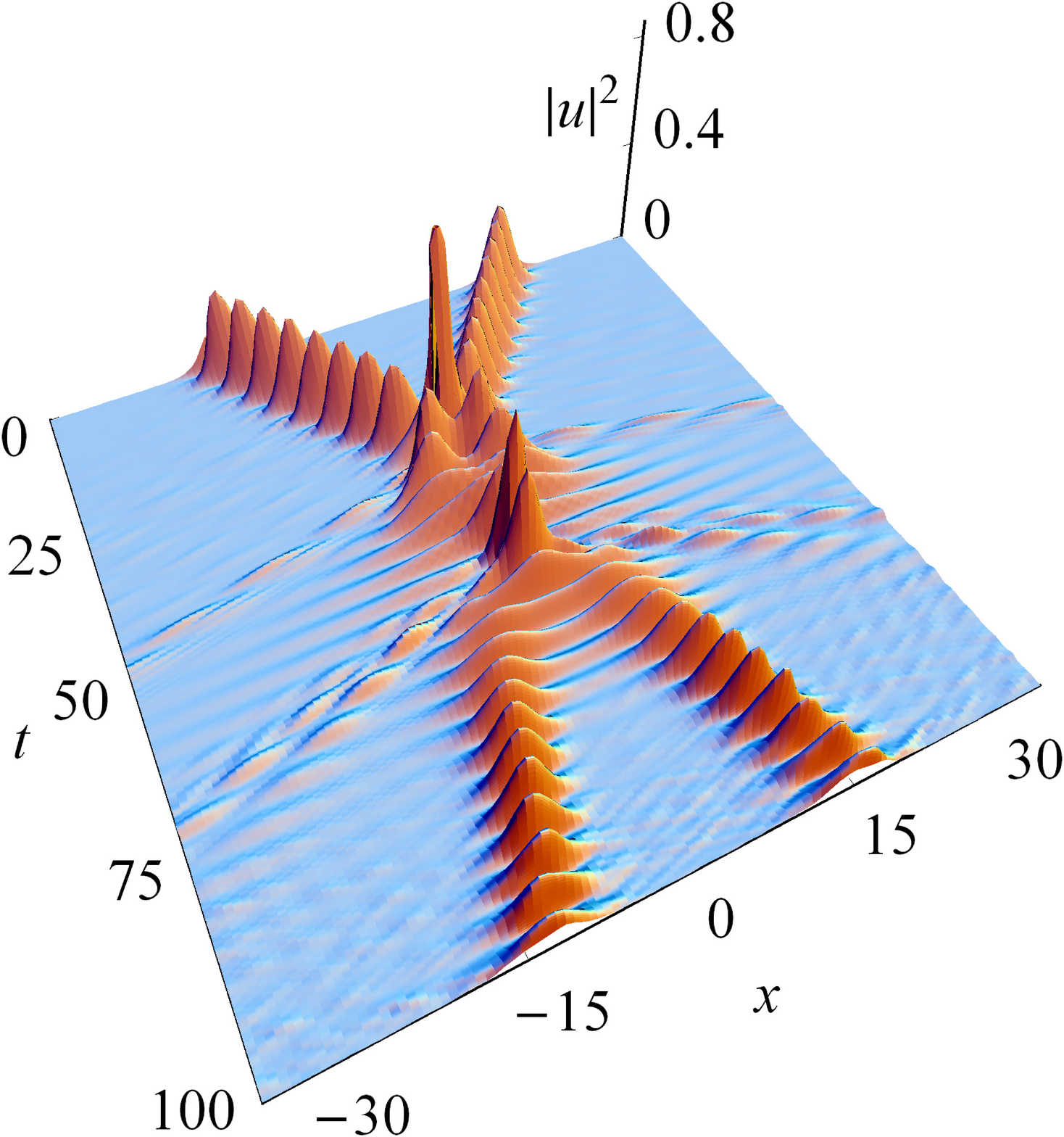}

  \caption{(Color online) The collision of two breathers.  Both breathers
  are taken  in the form
  \eqref{A110}, \eqref{A111}, \eqref{A105}, with the amplitudes
  $\sqrt{\epsilon}=0.3$, and Galilei-boosted  with the velocities  $v= \pm 0.5$.
  The panels (a) and (b) are different in the initial phase of the breathers.
    In both plots,
  $\gamma=0.3$.
  \label{bb}
 }
 \end{center}
 \end{figure}

Finally, we touch upon the collision
of two breathers.
The outcome of this collision
 can be predicted
on the basis of the amplitude equation \eqref{A21}.
Indeed,   the scattering  of two generic solitons in a  Hamiltonian system
 typically produces  two solitons of lower energy, or their bound state.
Consistently with these expectations, the numerical simulations
of Eqs.~\eqref{A1}
demonstrate the production of one or two breathers (Fig.~\ref{bb}).

  \section{Concluding remarks}
  \label{Conclusions}

  Stationary   solitons  in the ${\cal PT}$ symmetric
  planar coupler are known
   to be sustained due to the exact offsetting of
  the power gained in the active waveguide by the power lost
  in its passive counterpart \cite{Suchkov:2011-46609:PRE,Driben:2011-4323:OL,Alexeeva:2012-63837:PRA}. 
  In this paper, we have described another realisation of the gain-loss
  balance, which is provided by the breathers.
  In the breather case, the total power is conserved not
  at every moment in time, but only over a period of oscillation.

 Results of our study can be summarised as follows.

 1.  We have derived a system of amplitude equations [Eqs.~\eqref{A21}]  governing the envelope
 of the breather. For times $t \lesssim \epsilon^{-2}$, where $\epsilon^{1/2}$ gives the scale of the amplitude of  the
 small-amplitude breather, the system \eqref{A21} is equivalent to the original system \eqref{A1}.

2. Despite the fact that the original 
${\mathcal PT}$-symmetric system includes gain and loss, the amplitude system was
shown to be conservative.

 3. The breather solution was
 constructed as the asymptotic expansion \eqref{A110}, \eqref{A111}, \eqref{A105}.

 4.  We have proved that all small-amplitude  breathers  are stable
 on the timescale $t \lesssim \epsilon^{-2}$.
 The small-amplitude breather was shown to be  a ``simple" oscillation --- it cannot have any modulating frequencies in its spectrum.

 5.  Breathers were shown to be common occurrences in the ${\cal PT}$-symmetric chains
 of dimers. In particular, breathers are born in collisions of the low- and high-frequency
 solitons.

In conclusion, we need to make three remarks.
The first  one is on the ${\cal PT}$ breathers versus conservative
breathers and limit cycles.

   The ${\cal PT}$-symmetric breathers  are different from their conservative counterparts
   in that their associated physical observables (e.g. energy and momentum)
   are not stationary but oscillate in time.
   From this point of view,
   the ${\cal PT}$ breathers are similar to the time-periodic solitons
      in dissipative systems~\cite{Rosanov:2002:SpatialHysteresis,Akhmediev:2005:DissipativeSolitons,Alexeeva:1999-103:NLN, *Barashenkov:2002-104101:PRL, *Barashenkov:2011-56609:PRE}.
      However there is  an important distinction between the latter two categories
       too. Namely, the dissipative solitons are limit cycles
      (in an infinite-dimensional phase space);
      their amplitudes and periods are
      determined uniquely by the  parameters of the system.
     On the contrary, the ${\cal PT}$ breathers arise as members of two-parameter families,
     similar to periodic trajectories in Hamiltonian systems.

  \begin{figure}[t]
 \begin{center}
      \includegraphics*[width=\linewidth]{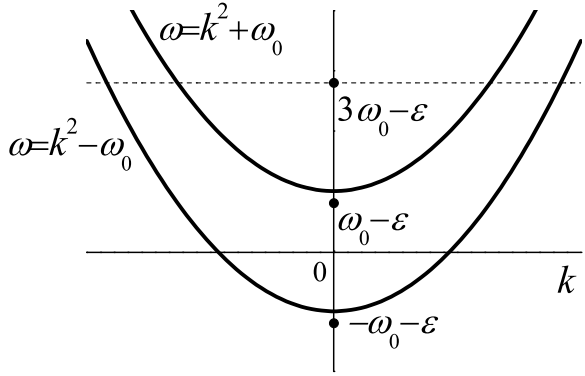}
       \caption{The dispersion curves of the ${\cal PT}$-symmetric
 system \eqref{A1}. The black dots indicate the two frequencies of the
 breather. The dashed line marks the frequency of radiation.
 \label{dispersion}}
 \end{center}
 \end{figure}

 The second remark is on the radiation from the breather.
 Using the singular perturbation expansion, the  breather can be constructed to
any order in $\epsilon$. All higher-order corrections $A_n$, $B_n$ are expressible as powers of $A_0$, $B_0$
and decay to zero as $|x| \to \infty$. There is no radiation to any order $\epsilon^n$, $n=0,1,2,3,...$.

However our simulations do reveal radiation waves from the breathers, with the amplitude of
waves growing as the amplitude of the breather is increased.
The reason why the asymptotic expansion does not capture these waves is
that the amplitude of radiation is exponentially small in $\epsilon$.
(The exponential smallness does not imply that the radiation is invisible for {\it finitely\/} small $\epsilon$
though.)

The frequency of the radiation can be determined on the basis
of standard considerations.
Indeed, the spectrum of linear excitations of the system \eqref{A30}
consists of two branches, $\omega = k^2 + \omega_0$
and  $\omega = k^2 - \omega_0$
[Fig.~\ref{dispersion}], while
the breather of the amplitude $\epsilon^{1/2}$ has two basic frequencies,
$\omega_0-\epsilon$ and
$-\omega_0-\epsilon$ [see Eq.~\eqref{A111}].
The term $a^2b^*$ in \eqref{A30} oscillates
at a combination frequency
 $3 \omega_0-\epsilon$ which
 falls in the linear spectrum.
 Hence the dominant frequency of the  resonant
  radiation will be $3 \omega_0-\epsilon$, as  indicated by the dashed line in Fig.~\ref{dispersion}.
  (Note that the frequency $\omega_0-\epsilon$
  does not resonate with the bottom branch since the $a$ and $b$
  modes are not coupled to the linear order.)

 Finally, we note that the breathers realise the  periodic light switching between the waveguides with gain and loss.
  Unlike oscillations in structureless   linear \cite{Guo:2009-93902:PRL,Ruter:2010-192:NPHYS} and nonlinear \cite{Sukhorukov:2010-43818:PRA,Ramezani:2010-43803:PRA}
  $\mathcal{PT}$ couplers,
  the breathers describe switching  between spatially extended waveguides.
  Here, the nonlinearity  suppresses the beam diffraction
  while the spatial coupling
   synchronizes the power oscillations across the beam.

\section*{Acknowledgements}

Useful conversations with Rodislav Driben, Sergey Flach,  Boris Malomed,  Dmitry Pelinovsky, and Dmitry Skryabin
are gratefully acknowledged. Special thanks go to Vladimir Konotop 
for his critical reading of the manuscript.
This work was supported by
the National Research Foundation
of South Africa
(grant UID  78952),
Russian Foundation for Fundamental Research
(grant 11-08-97057-p$\_$povoljie$\_$a), and
the Australian Research Council programs including Future Fellowship FT100100160.
 IB's work in Canberra was funded via the Visiting Fellowship of the
ANU.

\appendix

\section{More general breather solutions}
\label{MG}

In this Appendix, we briefly comment on other solutions
of the system \eqref{A22} --- more general than
a nearly symmetric configuration \eqref{G1}, \eqref{C1}, \eqref{P1Q1}.

By rescaling $P$, $Q$, $X$ and redefining $\ \epsilon$, we can always
arrange that $\mu=1$ in equations \eqref{A22}:
\begin{align}
  P''-  P  +2 (|P|^2+ 2 |Q|^2)P
+\frac{\epsilon}{\omega_0} (|Q|^4-2 |P Q|^2) P=0,  \nonumber
 \\
 Q'' -\nu Q +2 (|Q|^2+ 2 |P|^2)Q
 +\frac{ \epsilon}{ \omega_0} (2|P Q|^2-|P|^4)  Q=0. \label{A23}
\end{align}
Note that we are not setting $\nu$ equal to 1, along with $\mu$.

For $ \epsilon=0$, the system \eqref{A23} has  even and odd solutions
 with $n$ humps ($n=1, 2, ... $),
 with both $P$ and $Q$ being nonzero \cite{Haelterman:1994-3376:PRE,Yang:1997-92:PD}.
 Each of these can be used as a starting point in the regular perturbation expansion in powers of $\epsilon$.

 In particular, the  solution of the system \eqref{A23} with $ \epsilon=0$,
 with an even single-humped $P(X)$ and an even single-humped $Q(X)$,
   exists for
 $\alpha^{-2} < \nu < \alpha^2$, where $\alpha^2= \frac14 (\sqrt{17}-1)^2 \approx 2.438$, $\alpha^{-2} \approx 0.410$ \cite{Haelterman:1993-145:OC,Yang:1997-92:PD}.
 Therefore the system \eqref{A23} with sufficiently small {\it nonzero} $ \epsilon$ will also
 have a localised solution for any  $\nu$  between  $\alpha^{-2}$ and  $\alpha^2$.

 The solution with a two-humped even $P(X)$ and a two-humped odd $Q(X)$ exists for $\beta^2<\nu<1$, where
 $\beta^2=\frac14 (\sqrt{17}-3)^2 \approx 0.315$ \cite{Haelterman:1993-145:OC,Yang:1997-92:PD}.

 All these soliton-like solutions of the system \eqref{A22} give rise to breather solutions of
 the ${\cal PT}$-symmetric system \eqref{A1}.
 Thus for each $n \geq 1$, the system \eqref{A1} has a two-parameter family of nonpropagating
 breather solutions with $n$ humps.
 Representatives of the family are different in the amplitude and width of the humps, as well
 as the contrast of the $|u|^2$- and  $|v|^2$-oscillations.

\section{No internal modes  for the small-amplitude breather}
\label{IM}

The aim of this Appendix is  to show that the operator \eqref{H3} does not have
any discrete eigenvalues. To this end, we note that if $\lambda \neq 0$,
the bottom component of \eqref{H3} gives
\be
\int \rho_2(X) z^{(0)}(X) dX=0,
\label{H7}
\ee
where $z^{(0)}=\sech X$ is the null eigenvector of the operator $L_0$.
The constraint \eqref{H7} defines a subspace of the space of square integrable functions;
we will denote this subspace $\frak S$.

On the subspace $\frak S$, the operator $L_0$ is positive definite; hence we can write
\eqref{H3} as another scalar eigenvalue problem, alternative to \eqref{H4}:
\be
L_+ \rho_2 =  -\lambda^2 L_0^{-1} \rho_2, \quad
 \rho_2 \in \frak S.  \label{H10}
\ee

Assume the nonsymmetric matrix-differential operator in \eqref{H3} has
   nonzero eigenvalues $\lambda_1, \lambda_2, ...$.
  The corresponding eigenvalues $-\lambda_1^2< -\lambda_2^2< ...$ of \eqref{H10} are real, and the
   associated eigenfunctions $\rho_2(X)$ can also be chosen real.
   The lowest eigenvalue can be found
as the minimum of the Rayleigh quotient:
\be
-\lambda_1^2= \min_{\frak S}  \frac{(\rho_2, L_+ \rho_2)}{(\rho_2, L_0^{-1} \rho_2)}.
 \label{H6}
\ee
Since both $L_+$ and $L_0$ are positive definite,
Eq.~\eqref{H6} implies that
the eigenvalue $-\lambda_1^2$
 of the generalised eigenvalue problem \eqref{H10} is positive.
Hence $\lambda_1$  lies in the gap of the continuous spectrum of the
operator \eqref{H3}: $\lambda_1= i \omega_1$,
$-1< \omega_1<1$.

On the other hand, any function from $\frak S$ can be expanded over the continuous spectrum
eigenfunctions of the operator $L_0$:
\be
\rho_2(X)= \int {\mathcal R}(k) z_k(X) dk, \label{H8}
\ee
where $L_0 z_k=(1+2k^2) z_k$, $-\infty<k<\infty$.
Writing $L_+$ as $L_0+ \frac43 \sech^2  X$
and substituting \eqref{H8} in \eqref{H6}, the Rayleigh quotient becomes
\be
\frac{ \int {\mathcal R}^2(k)(1+2k^2) dk + \frac43 \int \rho_2^2 \sech^2 X dX}
{\int {\mathcal R}^2(k)(1+2k^2)^{-1} dk }.
\label{H9}
\ee
The first term in the numerator of \eqref{H9} is greater than the denominator;
hence the quotient  is greater than 1.
This contradicts the fact that the   eigenvalue
$\lambda_1$ is  in the gap of
the continuous spectrum of the operator \eqref{H3}.
The contradiction proves that the operator \eqref{H3} cannot have discrete eigenvalues.

\bibliographystyle{aps3AuAll}
\bibliography{db_BARASHENKOV}

\begin{thebibliography}{10}

\bibitem{Bender:1998-5243:PRL}
C.~M. Bender and S. Boettcher, Phys. Rev. Lett. {\bf 80}, 5243 (1998).

\bibitem{Bender:2007-947:RPP}
C.~M. Bender, Rep. Prog. Phys. {\bf 70}, 947 (2007).

\bibitem{Ruschhaupt:2005-L171:JPA}
A. Ruschhaupt, F. Delgado, and J.~G. Muga, J. Phys. A {\bf 38}, L171 (2005).

\bibitem{El-Ganainy:2007-2632:OL}
R. {G}anainy, {E}l, K.~G. Makris, D.~N. Christodoulides, and Z.~H. Musslimani,
  Opt. Lett. {\bf 32}, 2632 (2007).

\bibitem{Klaiman:2008-80402:PRL}
S. Klaiman, U. Guenther, and N. Moiseyev, Phys. Rev. Lett. {\bf 101}, 080402
  (2008).

\bibitem{Makris:2008-103904:PRL}
K.~G. Makris, R. {G}anainy, {E}l, D.~N. Christodoulides, and Z.~H. Musslimani,
  Phys. Rev. Lett. {\bf 100}, 103904 (2008).

\bibitem{Zheng:2010-10103:PRA}
M.~C. Zheng, D.~N. Christodoulides, R. Fleischmann, and T. Kottos, Phys. Rev. A
  {\bf 82}, 010103 (2010).

\bibitem{Berry:2008-244007:JPA}
M.~V. Berry, J. Phys. A {\bf 41}, 244007 (2008).

\bibitem{Longhi:2010-22102:PRA}
S. Longhi, Phys. Rev. A {\bf 81}, 022102 (2010).

\bibitem{Longhi:2009-123601:PRL}
S. Longhi, Phys. Rev. Lett. {\bf 103}, 123601 (2009).

\bibitem{Bendix:2009-30402:PRL}
O. Bendix, R. Fleischmann, T. Kottos, and B. Shapiro, Phys. Rev. Lett. {\bf
  103}, 030402 (2009).

\bibitem{Ruter:2010-192:NPHYS}
C.~E. Ruter, K.~G. Makris, R. {G}anainy, {E}l, D.~N. Christodoulides, M. Segev,
  and D. Kip, Nature Physics {\bf 6}, 192 (2010).

\bibitem{Guo:2009-93902:PRL}
A. Guo, G.~J. Salamo, D. Duchesne, R. Morandotti, M. {Volatier-Ravat}, V.
  Aimez, G.~A. Siviloglou, and D.~N. Christodoulides, Phys. Rev. Lett. {\bf
  103}, 093902 (2009).

\bibitem{Ramezani:2012-13818:PRA}
H. Ramezani, T. Kottos, V. Kovanis, and D.~N. Christodoulides, Phys. Rev. A
  {\bf 85}, 013818 (2012).

\bibitem{Miroshnichenko:2011-12123:PRA}
A.~E. Miroshnichenko, B.~A. Malomed, and {Yu}.~S. Kivshar, Phys. Rev. A {\bf
  84}, 012123 (2011).

\bibitem{West:2010-54102:PRL}
C.~T. West, T. Kottos, and T. Prosen, Phys. Rev. Lett. {\bf 104}, 054102
  (2010).

\bibitem{Sukhorukov:2012-2148:OL}
A.~A. Sukhorukov, S.~V. Dmitriev, S.~V. Suchkov, and {Yu}.~S. Kivshar, Opt.
  Lett. {\bf 37}, 2148 (2012).

\bibitem{Regensburger:2012-167:NAT}
A. Regensburger, C. Bersch, M.~A. Miri, G. Onishchukov, D.~N. Christodoulides,
  and U. Peschel, Nature {\bf 488}, 167–171 (2012).

\bibitem{Chen:1992-239:IQE}
Y.~J. Chen, A.~W. Snyder, and D.~N. Payne, IEEE J. Quantum Electron. {\bf 28},
  239 (1992).

\bibitem{Ramezani:2010-43803:PRA}
H. Ramezani, T. Kottos, R. {G}anainy, {E}l, and D.~N. Christodoulides, Phys.
  Rev. A {\bf 82}, 043803 (2010).

\bibitem{Sukhorukov:2010-43818:PRA}
A.~A. Sukhorukov, Z.~Y. Xu, and {Yu}.~S. Kivshar, Phys. Rev. A {\bf 82}, 043818
  (2010).

\bibitem{Musslimani:2008-30402:PRL}
Z.~H. Musslimani, K.~G. Makris, R. {G}anainy, {E}l, and D.~N. Christodoulides,
  Phys. Rev. Lett. {\bf 100}, 030402 (2008).

\bibitem{Dmitriev:2010-2976:OL}
S.~V. Dmitriev, A.~A. Sukhorukov, and {Yu}.~S. Kivshar, Opt. Lett. {\bf 35},
  2976 (2010).

\bibitem{Hu:2011-43818:PRA}
S.~M. Hu, X.~K. Ma, D.~Q. Lu, Z.~J. Yang, Y.~Z. Zheng, and W. Hu, Phys. Rev. A
  {\bf 84}, 043818 (2011).

\bibitem{Abdullaev:2011-41805:PRA}
F.~K. Abdullaev, Y.~V. Kartashov, V.~V. Konotop, and D.~A. Zezyulin, Phys. Rev.
  A {\bf 83}, 041805 (2011).

\bibitem{Shi:2011-53855:PRA}
Z.~W. Shi, X.~J. Jiang, X. Zhu, and H.~G. Li, Phys. Rev. A {\bf 84}, 053855
  (2011).

\bibitem{Zhu:2011-2680:OL}
X. Zhu, H. Wang, L.~X. Zheng, H.~G. Li, and Y.~J. He, Opt. Lett. {\bf 36}, 2680
  (2011).

\bibitem{Zezyulin:2011-64003:EPL}
D.~A. Zezyulin, Y.~V. Kartashov, and V.~V. Konotop, Europhys. Lett. {\bf 96},
  64003 (2011).

\bibitem{Nixon:2012-23822:PRA}
S. Nixon, L.~J. Ge, and J.~K. Yang, Phys. Rev. A {\bf 85}, 023822 (2012).

\bibitem{Hu:2012-43826:PRA}
S.~M. Hu, X.~K. Ma, D.~Q. Lu, Y.~Z. Zheng, and W. Hu, Phys. Rev. A {\bf 85},
  043826 (2012).

\bibitem{Zezyulin:2012-43840:PRA}
D.~A. Zezyulin and V.~V. Konotop, Phys. Rev. A {\bf 85}, 043840 (2012).

\bibitem{He:2012-3320:OC}
Y.~J. He, X. Zhu, D. Mihalache, J.~L. Liu, and Z.~X. Chen, Opt. Commun. {\bf
  285}, 3320 (2012).

\bibitem{Zeng:2012-47601:PRE}
J.~H. Zeng and Y.~H. Lan, Phys. Rev. E {\bf 85}, 047601 (2012).

\bibitem{Suchkov:2011-46609:PRE}
S.~V. Suchkov, B.~A. Malomed, S.~V. Dmitriev, and {Yu}.~S. Kivshar, Phys. Rev.
  E {\bf 84}, 046609 (2011).

\bibitem{Driben:2011-4323:OL}
R. Driben and B.~A. Malomed, Opt. Lett. {\bf 36}, 4323 (2011).

\bibitem{Driben:2011-51001:EPL}
R. Driben and B.~A. Malomed, Europhys. Lett. {\bf 96}, 51001 (2011).

\bibitem{Alexeeva:2012-63837:PRA}
N.~V. Alexeeva, I.~V. Barashenkov, A.~A. Sukhorukov, and {Yu}.~S. Kivshar,
  Phys. Rev. A {\bf 85}, 063837 (2012).

\bibitem{Kosevich:1974-1793:ZETF}
A.~M. Kosevich and A.~S. Kovalev, Zh. \'Eksp. Teor. Fiz. {\bf 67}, 1793 (1974)
  (in Russian) [JETP {\bf 67}, 891 (1975)].

\bibitem{Dashen:1975-3424:PRD}
R.~F. Dashen, B. Hasslacher, and A. Neveu, Phys. Rev. D {\bf 11}, 3424 (1975).

\bibitem{Segur:1987-747:PRL}
H. Segur and M.~D. Kruskal, Phys. Rev. Lett. {\bf 58}, 747 (1987).

\bibitem{Boyd:1990-177:NLN}
J.~P. Boyd, Nonlinearity {\bf 3}, 177 (1990).

\bibitem{Li:2011-66608:PRE}
K. Li and P.~G. Kevrekidis, Phys. Rev. E {\bf 83}, 066608 (2011).

\bibitem{Dmitriev:2011-13833:PRA}
S.~V. Dmitriev, S.~V. Suchkov, A.~A. Sukhorukov, and {Yu}.~S. Kivshar, Phys.
  Rev. A {\bf 84}, 013833 (2011).

\bibitem{Szameit:2011-21806:PRA}
A. Szameit, M.~C. Rechtsman, O. Bahat~Treidel, and M. Segev, Phys. Rev. A {\bf
  84}, 021806 (2011).

\bibitem{Suchkov:2012-33825:PRA}
S.~V. Suchkov, S.~V. Dmitriev, B.~A. Malomed, and {Yu}.~S. Kivshar, Phys. Rev.
  A {\bf 85}, 033825 (2012).

\bibitem{Kivshar:2003:OpticalSolitons}
Yu.~S. Kivshar and G.~P. Agrawal, {\em {Optical Solitons: From Fibers to
  Photonic Crystals}} (Academic Press, San Diego, 2003).

\bibitem{Ueda:1990-563:PRA}
T. Ueda and W.~L. Kath, Phys. Rev. A {\bf 42}, 563 (1990).

\bibitem{Malomed:1991-1388:OL}
B.~A. Malomed and S. Wabnitz, Opt. Lett. {\bf 16}, 1388 (1991).

\bibitem{Mesentsev:1992-1497:OL}
V.~K. Mesentsev and S.~K. Turitsyn, Opt. Lett. {\bf 17}, 1497 (1992).

\bibitem{Kaup:1993-3049:PRE}
D.~J. Kaup, B.~A. Malomed, and R.~S. Tasgal, Phys. Rev. E {\bf 48}, 3049
  (1993).

\bibitem{Haelterman:1994-3376:PRE}
M. Haelterman and A. Sheppard, Phys. Rev. E {\bf 49}, 3376 (1994).

\bibitem{Haelterman:1993-145:OC}
M. Haelterman, A.~P. Sheppard, and A.~W. Snyder, Opt. Commun. {\bf 103}, 145
  (1993).

\bibitem{Silberberg:1995-246:OL}
Y. Silberberg and Y. Barad, Opt. Lett. {\bf 20}, 246 (1995).

\bibitem{Yang:1996-111:STAM}
J. Yang and D.~J. Benney, Stud. Appl. Math. {\bf 96}, 111 (1996).

\bibitem{Yang:1997-92:PD}
J.~K. Yang, Physica D {\bf 108}, 92 (1997).

\bibitem{Yang:1997-61:STAM}
J.~K. Yang, Stud. Appl. Math. {\bf 98}, 61 (1997).

\bibitem{Yang:2001-26607:PRE}
J.~K. Yang, Phys. Rev. E {\bf 64}, 026607 (2001).

\bibitem{Tan:2001-56616:PRE}
Y. Tan and J.~K. Yang, Phys. Rev. E {\bf 64}, 056616 (2001).

\bibitem{Driben:2012:EPL}
R. Driben and B.~A. Malomed, Europhys. Lett. {\bf 99} (2012), in press;
  preprint arXiv:1207.3917.

\bibitem{Rosanov:2002:SpatialHysteresis}
N.~N. Rosanov, {\em Spatial Hysteresis and Optical Patterns}, {\em Springer
  Series in Synergetics} (Springer, New York, 2002).

\bibitem{Akhmediev:2005:DissipativeSolitons}
{\em Dissipative Solitons}, {\em Lecture Notes in Physics}, N. Akhmediev and A.
  Ankiewicz, eds., (Springer, New York, 2005).

\bibitem{Alexeeva:1999-103:NLN}
N.~V. Alexeeva, I.~V. Barashenkov, and D.~E. Pelinovsky, Nonlinearity {\bf 12},
  103 (1999).

\bibitem{Barashenkov:2002-104101:PRL}
I.~V. Barashenkov, N.~V. Alexeeva, and E.~V. Zemlyanaya, Phys. Rev. Lett. {\bf
  89}, 104101 (2002).

\bibitem{Barashenkov:2011-56609:PRE}
I.~V. Barashenkov, E.~V. Zemlyanaya, and T.~C. {van Heerden}, Phys. Rev. E {\bf
  83}, 056609 (2011).

\end{thebibliography}

\end{document}